\documentclass[fleqn,usenatbib]{mnras}
\usepackage{newtxtext,newtxmath}

\usepackage[T1]{fontenc}
\DeclareRobustCommand{\VAN}[3]{#2}
\let\VANthebibliography\thebibliography
\def\thebibliography{\DeclareRobustCommand{\VAN}[3]{##3}\VANthebibliography}

\usepackage{graphicx}
\usepackage{amsmath}
\usepackage{subcaption}
\usepackage{tabularx}
\usepackage{soul}
\usepackage{lipsum}

\newcommand{\orcid}[1]{\href{https://orcid.org/#1}{\includegraphics[scale=0.06]{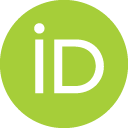}}}

\title[Inferring the ICM properties in TNG Cluster]{The temperature and metallicity distributions of the ICM: \\insights with TNG-Cluster for \textit{XRISM}-like observations}

\author[D. Chatzigiannakis et al.]{Dimitris Chatzigiannakis \orcid{0009-0008-3247-9489},$^{1,2}$\thanks{E-mail: chatzigiannakis@mpia.de}
Annalisa Pillepich \orcid{0000-0003-1065-9274},$^{1}$
Aurora Simionescu,\orcid{0000-0002-9714-3862}$^{3,4,5}$
Nhut Truong \orcid{0000-0003-4983-0462},$^{6,7}$\newauthor
and Dylan Nelson \orcid{0000-0001-8421-5890}$^{8}$
\\
$^{1}$Max-Planck-Institut für Astronomie, Königstuhl 17, D-69117 Heidelberg, Germany \\
$^{2}$Fakultät für Physik und Astronomie, Universität Heidelberg, Im Neuenheimer Feld 226, 69120 Heidelberg, Germany\\
$^{3}$SRON Netherlands Institute for Space Research, Niels Bohrweg 4, 2300 RA Leiden, The Netherlands\\
$^{4}$Leiden Observatory, Leiden University, Niels Bohrweg 2, 2300 RA Leiden, The Netherlands\\
$^{5}$Kavli Institute for the Physics and Mathematics of the Universe (WPI), The University of Tokyo, Kashiwa, Chiba 277-8583, Japan\\
$^6$\ Center for Space Sciences and Technology, University of Maryland, Baltimore County, 1000 Hilltop Circle, Baltimore, MD 21250, USA \\
$^{7}$ NASA Goddard Space Flight Center, Code 662, Greenbelt, MD 20771, USA\\
$^{8}$ Universität Heidelberg, Zentrum für Astronomie, ITA, Albert-Ueberle-Str. 2, 69120 Heidelberg, Germany
}
\date{}

\pubyear{2025}

\begin{document}
\label{firstpage}
\pagerange{\pageref{firstpage}--\pageref{lastpage}}
\maketitle

\begin{abstract}
The new era of high-resolution X-ray spectroscopy will significantly improve our understanding of the intra-cluster medium (ICM) by providing precise constraints on its underlying physical properties. However, spectral fitting requires reasonable assumptions on the thermal and chemical distributions of the gas. We use the output of TNG-Cluster, the newest addition to the IllustrisTNG suite of cosmological magnetohydrodynamical simulations, to provide theoretical expectations for the multi-phase nature of the ICM across hundreds of $z=0$ clusters (M$_{\rm{500c}} = 10^{14.0-15.3}~\rm{M}_\odot$) based upon a realistic model for galaxy formation and evolution. We create and analyse, in an observer-like manner, end-to-end \textit{XRISM}/Resolve mock observations towards cluster centres. We then systematically compare the intrinsic temperature and Fe abundance of the simulated gas with the inferred ones from spectral fitting via a variety of commonly used spectral-emission models. Our analysis suggests that models with a distribution of temperatures, better describe the broad thermal distributions of the ICM, as predicted by TNG-Cluster, but still incur biases in the inferred temperature of $0.5-2$ keV (16th-84th percentiles). However, all spectral-emission models systematically underestimate the Fe abundance of the central ICM by 0.12~Solar ($\sim$ 22 per cent), almost an order of magnitude higher than the abundance errors reported in the literature, primarily due to projection effects. Selecting only strong cool core clusters leads to minor improvements on inference quality, removing the majority of outliers but maintaining similar overall biases and cluster-to-cluster scatter.
\end{abstract}



\section{Introduction}

Galaxy clusters are excellent probes of the large-scale structure of the Universe due to their rarity, masses and preferential position at the knots of the cosmic web \citep[e.g.][]{Allen11}. Their extended gaseous atmospheres consisting of X-ray emitting plasma, known as the intra-cluster medium (ICM), play a key role in our understanding of the evolution of large structures, since processes such as mergers, gas accretion from the cosmic web and feedback all leave signatures in the distribution and properties of this gas.

Various X-ray telescopes (e.g. \textit{XMM-Newton}, \textit{Chandra}, \textit{Suzaku}, \textit{SRG/eROSITA}) have studied the ICM and its thermal and chemical properties across spatial and mass scales \citep[e.g.][and references therein]{Kravtsov12,Mernier18c,Walker19,Mernier23}. However, the low spectral resolution of the CCD detectors ($\sim$120~eV) so far limits the scope of X-ray analyses. For example, at CCD resolution, the dynamical state of the gas is hard to assess, as it lacks the spectral resolution for a direct measurement of gas motions from the Doppler shifts of spectral lines. Similarly, neighbouring emission lines may get blended, complicating the recovery of e.g. the abundances of various elemental species and the extent of the multi-phase thermal structure of the gas. As a result, many simplifying assumptions, such as coupling individual abundances to Iron (Fe) or assuming single temperature gas, have been routinely made.

In the current and upcoming new era of high-resolution X-ray spectroscopy, from \textit{Hitomi} to missions such as \textit{XRISM}\footnote{\url{https://www.xrism.jaxa.jp/en/}} and \textit{newAthena}, these issues can be addressed. For example, as shown for Perseus, \textit{Hitomi}'s spectral resolution of $\sim$ 5~eV revealed a wealth of previously-unresolved spectral lines of different species in the $1.8-9$~keV band \citep[][]{HitomiColab17}. \textit{XRISM} has further promised to break the degeneracies in the thermal structure of the gas by providing robust measurements of the various thermal components in clusters.

However, by overcoming many previous degeneracies, systematic uncertainties become a more prominent source of errors. Assumptions made during the analysis of the X-ray spectra -- such as the assumed shape of the temperature and metallicity distributions of the ICM or on the atomic libraries describing the emission lines, as well as unavoidable projection effects -- all impact inferences, and may lead to significant biases. This can be problematic, especially for science cases that are sensitive to the exact values of cluster properties, such as their metal budgets \citep[e.g.][and references therein]{Gastaldello21} and the X-ray observable-mass relations used for cosmological inference \citep[e.g. most recently][]{Ghirardini24}.

Unfortunately, the physical properties of the ICM are sufficiently complex that these issues cannot be assessed empirically. In this case, numerical simulations are useful tools. With the inclusion of ever more and more realistic physical processes, simulations provide an increasingly detailed view of the ICM \citep[e.g.][and references therein]{Nelson24}. A number of previous works have quantified biases on the observationally inferred properties of clusters by mocking the X-ray emission of simulated clusters and by focusing on past, current and future X-ray missions \citep[e.g.][and references therein]{Gardini04,Rasia06,Rasia08,Biffi18,Cucchetti18,Roncarelli18,Mernier20,Barnes21,ZuHone23b,Castellani24,Veronesi24,Zhang24,Sanders23}. However, no analysis has yet quantified the systematic effects due to specific parametric model choices, such as the ICM temperature distribution, when fitting high-resolution X-ray spectra.

In this work, we aim to assess the ability of spectral-emission models that are routinely adopted in the field to infer the mass-weighted temperature and Fe abundance of the ICM in cluster centres. Specifically, we attempt to quantify the systematic errors that might be introduced in said inferences due to the assumed  description of the ICM's temperature distribution, by simultaneously accounting for cluster-to-cluster variations. To do so, we use TNG-Cluster \citep[][]{Nelson24}, a recent extension of the IllustrisTNG simulations,\footnote{\url{www.tng-project.org}} which returns a rich and statistically-large sample of simulated galaxy clusters well suited to study the ICM. Based on the IllustrisTNG model \citep{Weinberger17, Pillepich18a} applied to a suite of zoom-in cosmological magneto-hydrodynamical simulations, TNG-Cluster not only creates galaxies at $z=0$ that are consistent with observations, but also -- and of relevance to this work --  produces the ICM in clusters with overall characteristics that are consistent with observations \citep{Nelson24}. 

It is thanks to the combination of numerical resolution and the adopted physical modelling that the output of TNG-Cluster is reasonably, if not quantitatively, realistic. This includes a mix of strong, weak and non cool cores \citep[][]{Lehle24,Lehle25}, realistic levels of gas velocity dispersion \citep{Truong24} and kinematics \citep{Ayromlou24}, a non-negligible amount of cool gas \citep[][]{Rohr24b,Staffehl25}, a diversity of radio relics \citep{Lee24}, and rich X-ray morphologies similar to observations \citep{Chadayammuri25}, including cold fronts, shocks, and cavities \citep{Truong24, Prunier24}. Additionally, the ICM returned by TNG-Cluster exhibits thermodynamical profiles across the 352 clusters -- density, temperature, entropy, and pressure -- that are in the ballpark of observational inferences \citep{Lehle24}. The distribution of cooling radii, derived from spectral fitting on both TNG-Cluster mocks and real Chandra observations, are quantitatively consistent \citep[][]{Prunier25}. And the metallicity profiles appear compatible with available observational results taken at face value \citep{Nelson24}. Overall, the realism of the ICM produced by TNG-Cluster further supports the credibility of the analysis we set out.

With TNG-Cluster, we can hence examine the performance of the most commonly used temperature and metallicity models to fit observed X-ray spectra, as we can retrieve the ``true'' ICM properties directly from the simulations. Importantly, we do so across a sample of a few hundred clusters at any given redshift, hence providing robust expectations across the diverse cluster population. To make our analysis timely, we focus on the types of observations currently being taken with \textit{XRISM}, specifically of the central regions of nearby galaxy clusters with the soft X-ray spectrometer Resolve, for which we develop an end-to-end mock and analysis pipeline that we apply to the simulated clusters.

This paper is arranged as follows. In Section~\ref{sec:input} we present our sample of simulated galaxy clusters from TNG-Cluster. In Section~\ref{sec:mock_proc} we describe our end-to-end mocking procedure, used to create \textit{XRISM}/Resolve like observations of our input clusters, and to analyse them. In Section~\ref{sec:intrinsic} we first examine the intrinsic properties of the ICM, as predicted by TNG-Cluster. We then compare the inferences of various spectral-emission modelling approaches on the temperature distribution and Fe abundance in Sections \ref{sec:temp} and \ref{sec:Fe_abund}, respectively. We discuss the reasons and implications of our findings and expand on various effects in Section~\ref{sec:discussion}, and summarize in Section~\ref{sec:conclusions}.

\section{Data and methods}

\subsection{The TNG-Cluster simulation suite}\label{sec:input}

TNG-Cluster is a collection of 352 zoom-in magnetohydrodynamical simulations of galaxy clusters with halo masses of log$_{10}(M_{\rm{200c}}/\rm{M}_\odot) = [14.2,15.3]$ i.e. log$_{10}(M_{\rm{500c}}/\rm{M}_\odot) = [14.0, 15.3]$. These are drawn from a cubic parent box of about 1 comoving Gpc per side. It complements the volume-limited TNG300 simulation \citep[][]{Springel18, Pillepich18b, Nelson18a, Naiman18, Marinacci18} with the highest halo masses. In particular, a subset of halos with log$_{10}M_{\rm{200c}}= 14.5 - 15.0$ at $z=0$ have been randomly chosen for re-simulation in TNG-Cluster in order to produce a uniform selection, and hence to compensate for the halo mass drop in TNG300, while simultaneously including all halos with log$_{10}(M_{\rm{200c}})>15$ from the same parent box. The volumes for re-simulations have been chosen such that the high-resolution zoom-in region around each cluster has a radius typically larger than two times the halo virial radius. For a more detailed description of the halo selection and model choices we refer to \citet{Nelson24}. Here we summarize additional salient aspects.

TNG-Cluster was simulated with the AREPO code \citep[][]{Springel10} at a resolution of 1.1$\times 10^{7}$~M$_\odot$ in gas/stellar target mass, matching that of the TNG300 fiducial run. It is fully based on the IllustrisTNG model \citep[][]{Weinberger17, Pillepich18a}. This includes a variety of processes that are needed to simulate galaxy formation and evolution, such as stellar formation and evolution, metal enrichment, radiative cooling/heating, seeding and growth of SMBH and both stellar and SMBH (thermal and kinetic) feedback. Specifically, the IllustrisTNG model produces stellar particles, each representing a single-age stellar population with an initial mass function described by \citet{Chabrier03}. Such populations evolve and return metals to the surrounding inter-stellar medium via AGB stars, type Ia and type II supernovae over time \citep[][]{Vogelsberger13,Wiersma09} based on the stellar yields described in detail in \citet{Pillepich18a}. The simulation directly tracks nine elements (H, He, C, N, O, Ne, Mg, Si, and Fe), in addition to Europium, tagging their production channel.

$\Lambda$CDM cosmological parameters from the Planck observations of the cosmic microwave background \citep[][]{PlanckColab16}  are assumed in TNG-Cluster and throughout this paper: $\Omega_{\rm{m}}$=0.3089, $\Omega_\Lambda$=0.6911, $\Omega_{\rm{b}}$=0.0486, H$_0$=67.74~km~s$^{-1}$~Mpc~$^{-1}$, $\sigma_8$=0.8159 and n$_s$=0.9667, consistent with the other IllustrisTNG runs.

\begin{figure*}
    \centering
    \includegraphics[width=\textwidth]{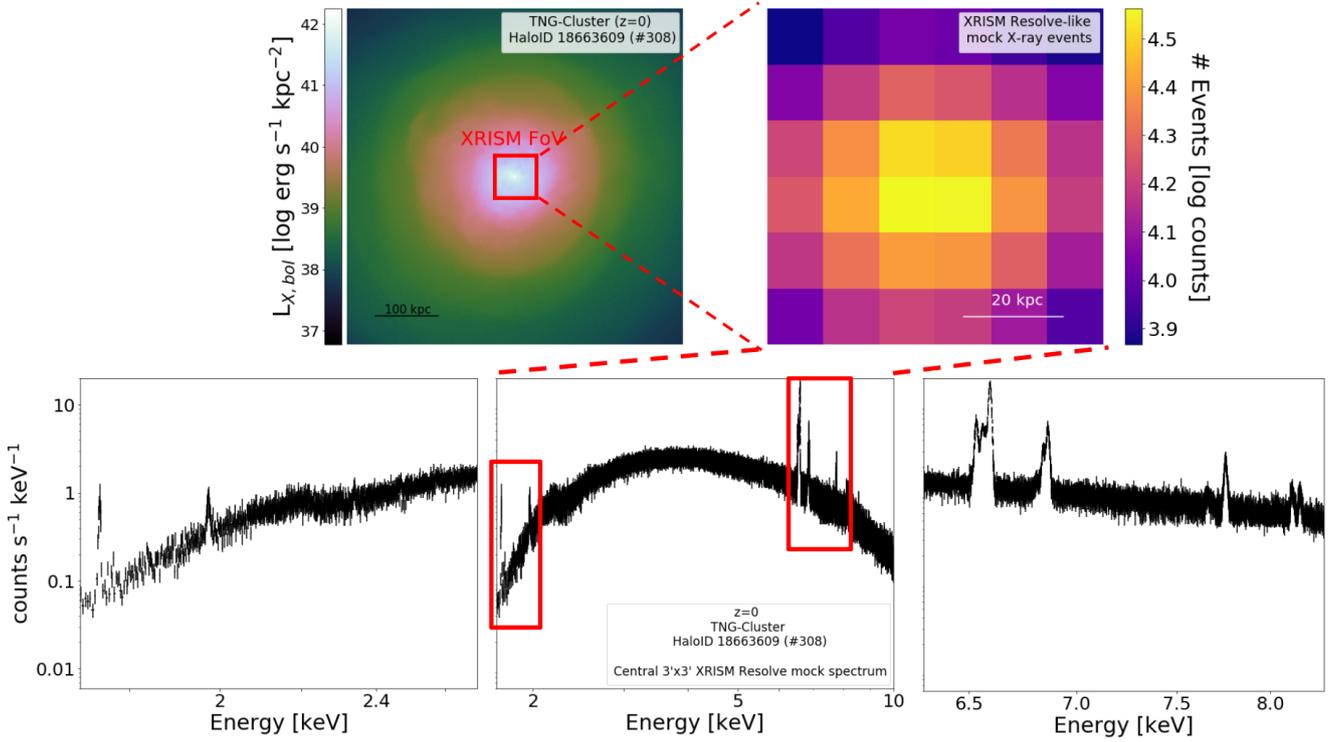}
    \caption{\textbf{Illustration of the end-to-end \textit{XRISM}/Resolve mocking procedure of simulated clusters from the TNG-Cluster suite.} On the top left, we present the intrinsic X-ray luminosity of a randomly-selected cluster from TNG-Cluster, highlighting the FoV of \textit{XRISM}/Resolve, pointing at the centre of the cluster. Following the process described in Section~\ref{sec:mock_proc}, we generate a mock event file as if it was observed by \textit{XRISM}/Resolve, presented in the top right corner. The mock spectrum extracted from the event file and a zoom-in on the Fe-K lines are presented in the bottom row. This process showcases our ability in generating high-quality mock spectra from our cosmologically-simulated clusters as if they were observed by \textit{XRISM}/Resolve.}
    \label{fig:xrism_mock}
\end{figure*}

\subsection{End-to-end \textit{XRISM}/Resolve-like mocks of simulated ICM}{\label{sec:mock_proc}}

In order to quantify the accuracy of the most commonly used X-ray emission models, we generate mock observations of our simulated clusters. In the following sections, we hence describe our end-to-end pipeline for \textit{XRISM}/Resolve-like X-ray mock spectra using TNG-Cluster halos.

\subsubsection{Modeling the intrinsic X-ray signal}
First, we model the X-ray emission of the ICM based on the thermodynamical and chemical properties of the simulation gas cells, as predicted by TNG-Cluster. For the purposes of this work, we use the 4.4.0 version of the PyXSIM package \citep[][]{ZuHone14}, which follows the approach of the PHOX code \citep[][]{Biffi12,Biffi13}, in combination to the SOXS package \citep[][]{ZuHone23}.

Similar to recent works \citep[e.g.][]{ZuHone23b,Truong24}, for every cluster, we consider all the gas cells that are within a 3D radius of $R_{\text{500c}}$, including contribution from satellites, as emitting sources. While this approach excludes emission from structures outside that radius, \citet{ZuHone23b} has shown for \textit{eROSITA} mocks that their effect on the inferred cluster properties is negligible. Throughout, the centre of a cluster from TNG-Cluster is defined as the minimum of the potential. Furthermore, we assume that all gas cells satisfying the following criteria contribute to the X-ray emission of a given targeted cluster: have a temperature higher than $3 \times 10^5$\,K ($\simeq$ 0.03 keV), a negative net cooling rate, and are not star forming. 

For each gas cell, we assume an optically-thin plasma in collisional ionisation equilibrium. We therefore model its emitted spectrum with an \texttt{apec} model (i.e. equivalently the \texttt{vapec} variant among XSPEC's models) based on the AtomDB atomic library \citep[v3.0.9][]{Smith01,Foster12}. 

The properties of the gas cells predicted within the TNG-Cluster simulation and relevant for this pipeline are: density, temperature, and absolute elemental abundances (w.r.t. Hydrogen). More specifically, for the latter, we use the TNG-Cluster outcome for all nine elements that the simulation tracks and a relative abundance ratio of $\rm{X/Fe}=1$ for the rest \citep[adopting][]{Asplund09}, assuming that the absolute Fe abundance is a reasonable proxy of the underlying metal enrichment and that solar abundance ratios hold for those elements. Additionally, we account for the 3D position and velocity of each gas cell, to properly account for their effects to the properties of the generated X-ray photons. Namely, effects such as thermal broadening and Doppler shift of the emission lines are explicitly modelled by PyXSIM, producing an accurate representation of the X-ray emission from the ICM of our simulated galaxy clusters.

\subsubsection{Observational strategy}\label{sec:mocks}

In order to generate \textit{XRISM}/Resolve mock event files, namely a list of detected photons of a given energy, from the simulated photons, we first project our halos along the z direction of the simulation box, representing a random direction for each cluster, and place them at the redshift of the Perseus Cluster ($z=0.0178$), as in \citet{Truong24}. Photoelectric absorption is also applied to the simulated photons, approximating with a \texttt{wabs} model the galactic absorption in the position of the Perseus cluster in the sky with a column density of $N_{\rm{H}} = 1.38 \times 10^{21} \rm{cm}^{-2}$ \citep{Kalberla05}. 

Every cluster is observed with a single pointing aimed at each cluster's centre and hence detecting the emission from all gas cells in a $3'\times3'\times2R_{500c}$ box volume: this hence represents a subset of  the original $\gtrsim3R_{200c}$ ($\gtrsim$ 4-5~pMpc) re-simulation volume of the zoom simulations and is meant to replicate the \textit{XRISM}/Resolve's Field of View (FoV). In order to best approximate realistic observations of our targets, we also apply \textit{XRISM}/Resolve's instrumental responses to the intrinsic incoming photons. Namely, we use the Cycle 1 response files, as they have been incorporated in SOXS, for a high resolution ($\sim$ 5eV) response matrix file (RMF) and a standard point-source ancillary response file (ARF) with the gate valve closed. 

Regarding the background, as no instrumental background file is publicly available, we ignore it. On the other hand, for the foreground/background signals from the sky, we estimate that resolved point sources, the Cosmic X-ray Background (CXB), thermal emission from the Milky Way (MW) and the Local Hot Bubble (LHB) all provide a statistically-insignificant amount of detected photons and are effectively ignored in our mocks. This is a direct result of the small detector FoV, and its low efficiency in detecting low-energy photons due to the closed gate valve -- their contribution is more relevant in the lower energy regime. 

As a result of the procedure above, we obtain mock event files of \textit{XRISM}/Resolve-like observations centred at the cores, unless otherwise specified, of the 352 clusters at $z=0$ from TNG-Cluster, assuming a reasonable exposure time of 100ks for each system. 

Finally, we extract spectra from our event files using the \texttt{dmcopy} command from the \textit{Chandra} Interactive Analysis of Observations (CIAO v.4.16) assuming a default spectral extraction, for the entirety of the FoV of the mock observation. A schematic representation of the result of our mock-observation pipeline can be seen in \autoref{fig:xrism_mock}.

\subsection{Analysis i.e. fits of \textit{XRISM}/Resolve-like spectra} 
\label{sec:mock_analysis}

For the purposes of our analysis, we will be using the X-ray spectral-fitting program \texttt{XSPEC} \citep{Arnaud96}, using the AtomDB atomic database (v3.0.9), consistent with our X-ray mocks. 

Throughout this work, we utilize four commonly-used spectral-emission models with different assumptions on the thermal structure i.e. temperature distribution of gas. The models in question are:
\begin{itemize}
    \item \texttt{bvapec}, as the baseline model commonly used to describe the gas: this assumes that the observed ICM is described by a single temperature, i.e. it returns a single-point estimate for its temperature;
    \item double \texttt{bvapec}: this consists of two \texttt{bvapec} models and is primarily used to describe two distinct phases of gas (hot and cooler) assuming de facto a bimodal temperature distribution of the ICM;
    \item \texttt{bvgadem}, which assumes a Normal distribution of ICM temperatures around a central peak;
    \item \texttt{bvlognorm}\footnote{\url{https://github.com/jeremysanders/lognorm}}, which assumes a Log-Normal distribution of ICM temperatures around a central peak.
\end{itemize}

For the Fe abundance, all adopted models assume that the observed ICM is described by a single value, as is the case for temperature with the \texttt{bvapec} and double \texttt{bvapec} models. We note that for all cases, both the line broadening due to the kinematics of the gas and the abundances of Si, S, Ar, Ca and Ni are taken into consideration, are modelled independently, and are simultaneously fit for. This is motivated by the fact that, if left fixed at a specific values, degeneracies between parameters could influence the model inferences on the temperature and Fe abundance. Consistently with \citet{Truong24} and \citet{XrismCollab25d}, our inferred velocity dispersions are comparable to the values obtained by central \textit{XRISM} pointing observations (see \autoref{fig:vel_disp}). However, further analysis and their exploitation are part of a future work. We note that H/He is kept fixed at 1 Solar and the abundances of all remaining elements are coupled to Fe, since they don't have emission lines in our energy range.

For the purposes of this analysis, we do not deproject our spectra. With this choice we remain consistent with the spectral analysis of most recent \textit{XRISM} observations \citep[e.g.][]{XrismCollab25b,XrismCollab25c,Rose25,Fujita25}. The only exception is that of \citet{XrismCollab25a} for the Centaurus cluster, where a deprojected profile has been used for the interpretation of the results and has been constructed by fitting previous \textit{Chandra} observations of Centaurus. An exact replication of this particular approach and study would require the full mocking and spectral fitting of \textit{Chandra}-like observations for our entire sample, which would be beyond the scope of this project. Therefore, we choose to not account for projection effects on our spectral fitting, but we expand more on the implications of this choice in Section~\ref{sec:proj_eff}.

Regarding our spectral fitting strategy, all our extracted spectra are fitted in the [1.8-10]~keV range. Even though the target redshift is known, we allow for the fit to vary it over a $\pm$50 per cent range in order to account for possible bulk motions of the gas along our line of sight. We use a guided approach to fit the model parameters gradually. However, it is still possible for the model to produce a non-valid fit. For this reason we introduce convergence criteria.

\subsubsection{Acceptance criteria for spectral best fits}\label{sec:accept}

We distinguish between cases where the X-ray spectral fitting is successful (``accepted'') or not (''rejected''). The criteria are as follows.
 
For both the \texttt{bvgadem} and \texttt{bvlognorm} models, we require that the uncertainties on the central temperature and on the width of the distributions are smaller than the best-fit values themselves, as a minimum requirement for convergence. For the \texttt{bvgadem} model, we also require that the width of the assumed distribution is smaller than the best-fit central temperature, as otherwise the model would either imply the presence of negative temperatures or, due to its hard limit at 0, significantly alter its shape, in order to describe the thermal state of the gas. We note that this criterion is not as significant as the convergence of individual best-fit parameters, with $\sim$ 1 per cent of our clusters being rejected due to it. For the double \texttt{bvapec} model, we apply this same convergence criteria to each of the two temperature components, in order for the model to be meaningfully different from the single \texttt{bvapec} one. Lastly, we also apply a goodness-of-fit criterion to our spectra, requiring the C-statistic \citep[][]{Cash79} of our spectral fits to deviate less than 3$\sigma$ from its expected Cash-statistic C value, following \citet{Kaastra17}. If a best-fit result fails any of the above criteria, the fit is considered rejected; if it fulfils all of them, it is accepted.


\section{Results}

\begin{figure*}
\centering
        \includegraphics[width=1\textwidth]{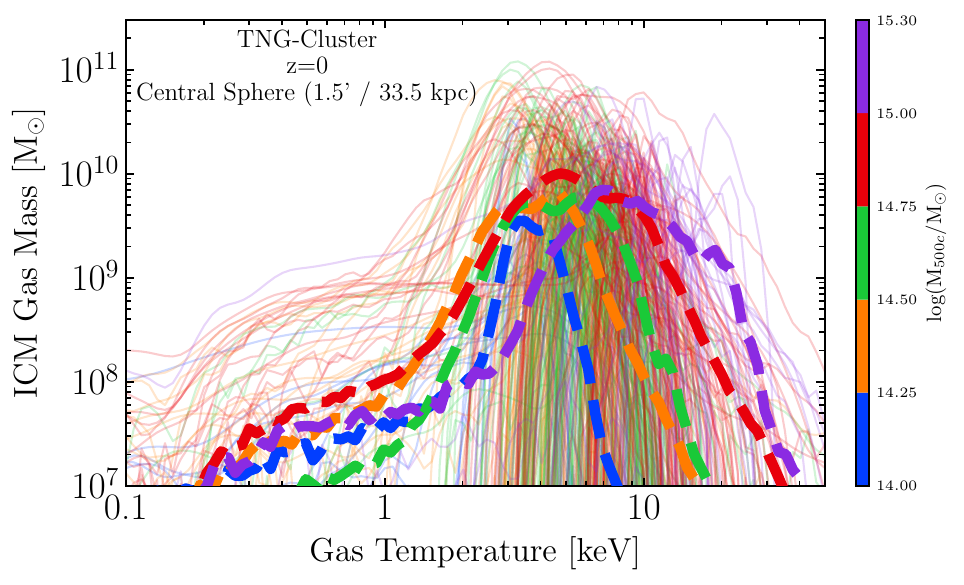}
        \includegraphics[width=1\textwidth]{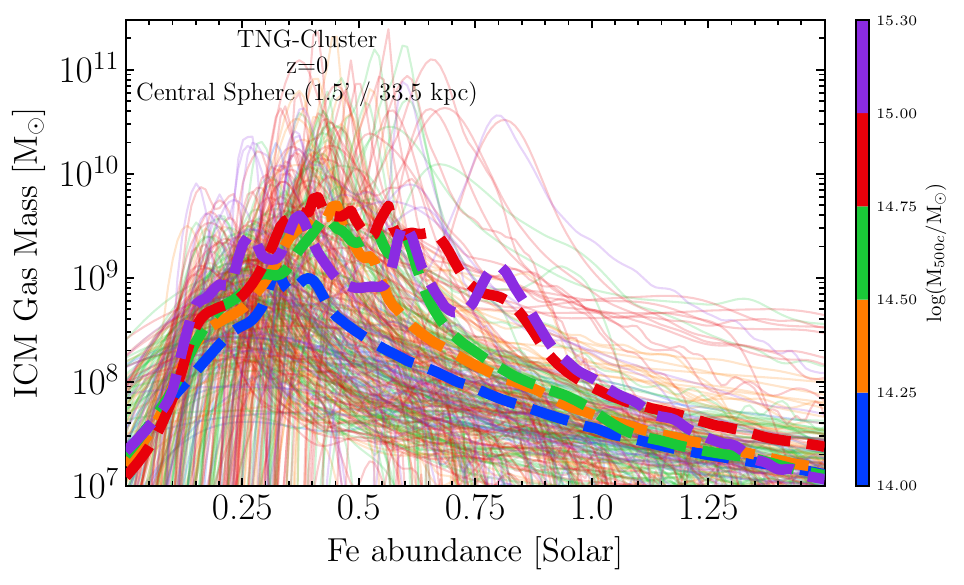}
    \caption{\textbf{Distributions of the gas temperature (top) and Fe abundance (bottom) for the central ICM of TNG-Cluster halos at z=0.} For both cases, each thin solid curve represents the mass-weighted Kernel Density Estimation (KDE) of all the gas cells inside the central 33.5~kpc (1.5' at the redshift of Perseus, to approximate \textit{XRISM} FoV) of a given simulated cluster, colour coded based on the M$_{\rm{500c}}$ halo mass. The thicker dashed curves indicate the running averages per halo mass bin, as indicated. TNG-Cluster (based on the IllustrisTNG model) predicts extended i.e. broad thermal and Fe abundance distributions for the ICM at the centre of clusters, with a singular peak, but with large variations from cluster to cluster, even for clusters with similar halo masses.}
    \label{fig:gas_prop}
\end{figure*}

\begin{figure*}
\centering
    \includegraphics[width=0.85\linewidth]{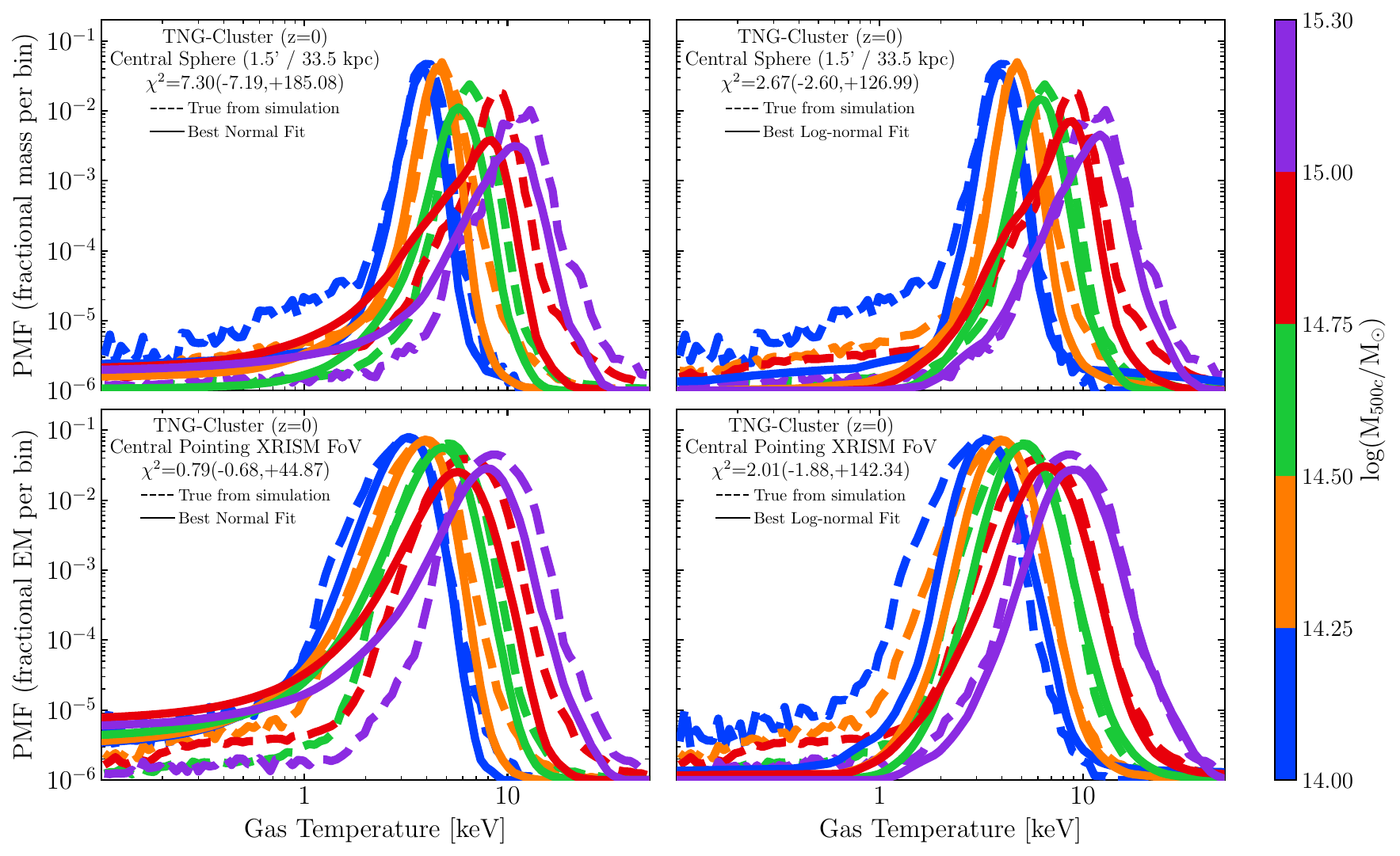}
    \caption{\textbf{What is the best functional form to describe the temperature distribution of the central ICM of TNG-Cluster?} The dashed curves represent the geometric mean of the mass-weighted gas in the 3D clusters' centres (top, as in \autoref{fig:gas_prop}) and of the emission-weighted gas for a central pointing with a \textit{XRISM}-like FoV (bottom), binned with respect to M$_{500\mathrm{c}}$ halo mass. More specifically, we plot the Probability Mass Functions (PMFs) normalized by the mass (top) and the Emission Measure (bottom; see Eq.~\ref{eq:EM}) of the gas in each bin. The solid curves represent the results of fitting the temperature distribution of each individual cluster with a Normal (left) and Log-Normal (right) functional form, then deriving the geometric mean across all cluster fits in each halo mass bin. We quote the median $\chi^2$ of the fits for our entire cluster sample, as well as its 16$^{\rm th}$ and 84$^{\rm th}$ percentile ranges. Residuals between the simulated and fitted temperature distributions of each studied cluster are provided in \autoref{fig:residuals}. Our analysis indicates that both functional forms may introduce systematic biases for the the description of the temperature distributions predicted by TNG-Cluster. Namely, the Normal distribution tends to overestimate the amount of cooler gas present in the ICM, while the Log-Normal distribution remains insensitive to the lower-temperature tails. The reported $\chi^2$ values would suggest that either model lacks the ability to fully capture the complexity of the simulations thermal distributions but both do a good job at describing the bulk of the ICM mass around the virial temperature peaks.}
    \label{fig:gas_prop_fit}
\end{figure*}

In the following sections we compare the ability of different spectral-fitting approaches to return the actual properties of the ICM, as predicted by TNG Cluster. We proceed with a multi-tiered approach.

Firstly, we compare the output temperature and Fe abundance of the best-fit spectral-emission models to the mass-weighted temperature and Fe abundance of the gas in the simulation located in a spherical volume centred at the centre of the cluster, with a radius of 1.5' (33.5~kpc at the redshift of Perseus, to approximate \textit{XRISM} FoV): we refer to this as gas in the central sphere or in the 3D cluster centre. Our choice is motivated by the fact that, observationally, one would hope that the output of the extracted spectrum represents the local conditions of the de-projected gas in the region of interest -- for us, the \textit{XRISM} FoV at the cluster centre. 

Secondly, we compare the spectral-emission model inferences to the emission-weighted properties of all the gas in the adopted FoV for a central pointing, namely also considering all the foreground and background ICM gas: we refer to this as gas in a central pointing with a XRISM-like FoV. Previous studies have shown that the observed ICM abundance profiles can be better reproduced by weighting the properties of gas particles/cells with emission rather than mass \citep[][]{Biffi17,Biffi18}. Therefore, this can be considered a more apt comparison with our observables and a characterization of the ``true'' ICM state that is by construction closer to what the spectral fitting can return.

For the purposes of this analysis, we use the emission measure (EM) of each gas cell as its weight:
\begin{equation}
    \text{EM}=\int n_\text{e} n_\text{H} dV \simeq n_\text{e} n_\text{H} V
    \label{eq:EM}
\end{equation}
where $n_\text{e}$, $n_\text{H}$ are the electron and hydrogen number densities and $V$ the volume of the gas cell, tracked by the simulation. We note that this weighting scheme differs from emission-weighted schemes that use the emissivity of each gas cell as their weight. Our choice in the weighting scheme is motivated by the fact that EM is indicative of each gas cell's contribution to the spectrum, as the basis of the spectral-emission model's normalisation. Furthermore, \citet{Cucchetti18} have indicated that the metallicity of the gas, weighted by its EM, can be accurately recovered by fitting mock, high-resolution spectra.

\subsection{The complexity of the thermal and chemical structure of cluster centres as predicted by TNG-Cluster}\label{sec:intrinsic}

We first quantify the properties of the central X-ray emitting gas, in the ICM of our 352 targets from the TNG-Cluster simulation.\\

{\it Temperature.} Starting with the temperature distribution of the gas in the centre of our clusters (\autoref{fig:gas_prop}, top), we find that, according to TNG-Cluster, the majority of the ICM mass is concentrated around a single temperature, the virial temperature of the cluster. The location of the peak exhibits, as expected, a strong correlation with halo mass, with the central ICM of more massive clusters being, on average, hotter. The width of the temperature distributions varies significantly across clusters, even between systems with similar halo masses. A subset of systems, independent of halo mass, appear to have extended tails of gas, extending to sub-keV temperatures. Considering that we are looking at the centrally-located gas, it is very likely that those tails are associated with the cooler inter-stellar (ISM) or circum-galactic (CGM) media of the brightest cluster galaxy \citep[BCG; see also][]{Rohr24b}. A few clusters also exhibit very high, super-virial temperature tails, which, given the SMBH feedback model implemented in TNG-Cluster, may be driven by central energy injections \citep{Truong24}. Overall, by visual inspection of the temperature distribution of each individual simulated cluster, we identify that at least 20 per cent of them (i.e. about 70 clusters of the 352) exhibit ICM in the central regions extending across more than one order of magnitude in temperature (in units of keV), i.e. with extended tails either towards cooler or super-virial temperatures.

Both the width and shape of the temperature distribution are relevant aspects for our analysis. So, what is the functional form of the temperature of the ICM at the centres of clusters favoured by TNG-Cluster? 

\begin{figure}
\centering
        \includegraphics[width=\columnwidth]{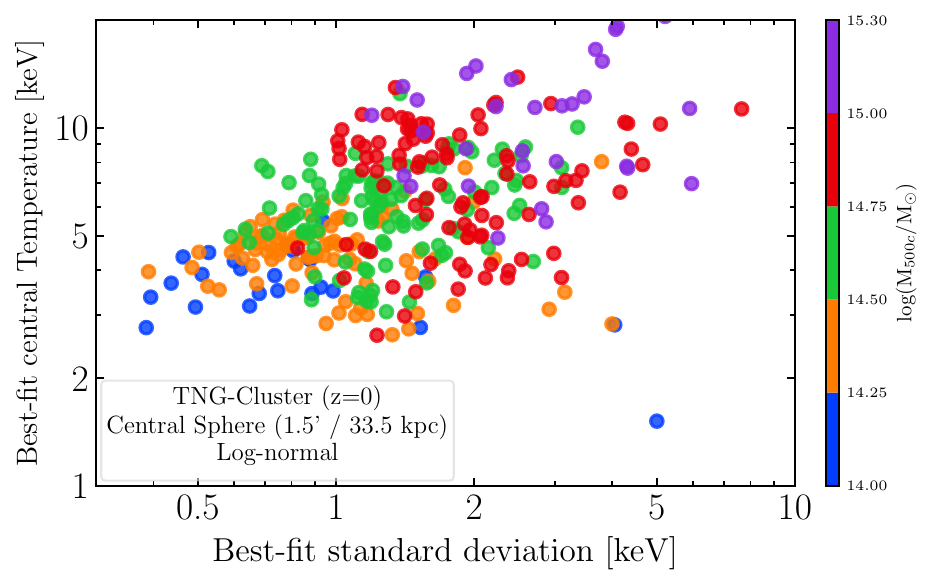}
    \caption{\textbf{Best-fit parameters of the Log-Normal distribution adopted to describe the ICM temperature of clusters' centres from TNG-Cluster.} We present the central temperature and standard deviation of the best-fit Log-Normal distributions for every TNG-Cluster halo at $z=0$, colour coded by its M$_{\rm{500c}}$ halo mass. The average distributions for each mass bin are shown in \autoref{fig:gas_prop} and \autoref{fig:gas_prop_fit}. While there is an obvious correlation between the best-fit central temperature and the halo mass, no such strong correlation exists for the standard deviation: how broad the temperature distribution is does not depend on total mass and can vary by an order of magnitude depending on the individual systems.}
    \label{fig:best_fit_val}
\end{figure}

We fit the temperature distributions in the 0.1-50~keV range with both the Normal and Log-Normal distributions. In \autoref{fig:gas_prop_fit}, we compare the geometric mean and best-fit temperature distributions of clusters in halo mass bins, for the intrinsic, mass-weighted gas in the 3D cluster centres (top) and for the emission-weighted gas in a \textit{XRISM}-like FoV (bottom). We note that the average curves are not representative of each individual cluster's best fit, but rather are meant to highlight possible systematic trends with cluster mass. The corresponding best-fit values of the Log-Normal distribution for each cluster are shown in \autoref{fig:best_fit_val}. The quality of each fit on a cluster-by-cluster basis is provided by the residuals in \autoref{fig:residuals}, and it is summarized by the median $\chi^2$ across the entire cluster sample annotated as text in \autoref{fig:gas_prop_fit}.

As \autoref{fig:gas_prop_fit} suggests, the temperature distribution cannot be perfectly described by either a Normal or a Log-Normal distribution in all systems. Both functional forms appear to be a reasonably good descriptor of the bulk of the temperature distribution around the virial temperature peaks. However, both models introduce biases primarily due to the presence of extended tails of gas in either direction, i.e. in those clusters where such tails are present. Both for the mass-weighted, centrally located gas as well as for the emission-weighted gas along the line of sight, the best-fit Normal distribution may underestimate to some small degree the amount of virial and hotter gas while typically systematically overestimating the amount of cooler gas, compared to the mode, present in the ICM.\footnote{Whereas the biases in the fits appear more pronounced in the most massive halos, this is an artifact of the averaging across multiple single-cluster distributions.}

On the other hand, the Log-Normal model appears to be totally insensitive to the extended tails of sub-keV gas, which it fails to capture. This effect is also reflected in the $\chi^2$, where, for both gas selections, the model tends to under-fit the true distribution of temperatures. Amid this overall picture, the cluster-to-cluster variation remains significant in all cases, as can be appreciated in \autoref{fig:residuals}. From the visual inspection of the fits to each cluster temperature distribution, the Log-Normal model also tends to grossly underestimate the amount of gas at the hottest-end of the temperature distribution (in about 100 clusters of 352).

Overall, our analysis suggests that, a priori, neither functional form provides a perfectly accurate representation of the underlying thermal properties of the gas for all clusters. While both the Normal and Log-normal distributions are certainly more informative than a single or double delta function, they can not fully capture the tails towards cooler and hotter gas present in the intrinsic temperature distributions of a non-negligible fraction of clusters. It is evident that more sophisticated models are needed to fully capture the thermal structure of the ICM, but a more in depth analysis and the development of such models is outside the scope of this paper.

In \autoref{fig:best_fit_val} we provide the best-fit central temperature and standard deviation for the Log-Normal fits, colour coded by the total mass of the cluster. Whereas there is a clear, and known, trend between average central ICM temperature and total cluster mass (spanning from a few keV to more than 10 keV for the studied clusters with log$_{10}M_{\rm{500c}}= 14.0 - 15.3$ at $z=0$), the ICM in each cluster exhibits a diverse range of temperatures, irrespective of total mass. According to TNG-Cluster, the Log-Normal width of the central ICM temperature distribution varies between 0.4 and 10 keV. \\

{\it Fe abundance.} Regarding the Fe abundance of the ICM in the centres of our simulated clusters (\autoref{fig:gas_prop}, bottom), TNG-Cluster predicts large system-to-system variations and broad distributions within individual clusters, across more than one order of magnitude in Solar units.

Upon inspection of all individual systems, the distributions can be broadly described as uni-modal, with the majority of the mass centred around a central peak for the largest majority of the clusters. In fact, we confirm that only a handful of metal-rich clusters exhibit secondary peaks at higher Fe abundances. However, unlike the temperature distributions, the peaks of the Fe abundance distribution show little to no correlation with halo mass: namely, clusters within the same halo mass bins can have significantly different amounts of Fe in their centres. We argue that this is related to their assembly and SMBH feedback histories, and we will expand on this in future works. Indeed, with the TNG model, simulated clusters have been shown to exhibit different central metallicity depending on their cool-core status \citep{Vogelsberger18}, akin to what is suggested by observations. Finally, similar to the tails of cooler gas found in the temperature distributions, a lot of systems tend to have extended tails towards super-solar Fe abundances from gas cells that appear to be associated with the ISM or CGM of the BCG.

Overall, given the broadness of the Fe abundances distributions predicted by TNG-Cluster, the single-metallicity models typically assumed for fitting X-ray spectra are a poor approximation of the metal-enrichment complexity of the centres of clusters.

\subsection{Inferring the ICM temperature distribution}\label{sec:temp}

\begin{figure*}
    \centering
    \includegraphics[width=\linewidth]{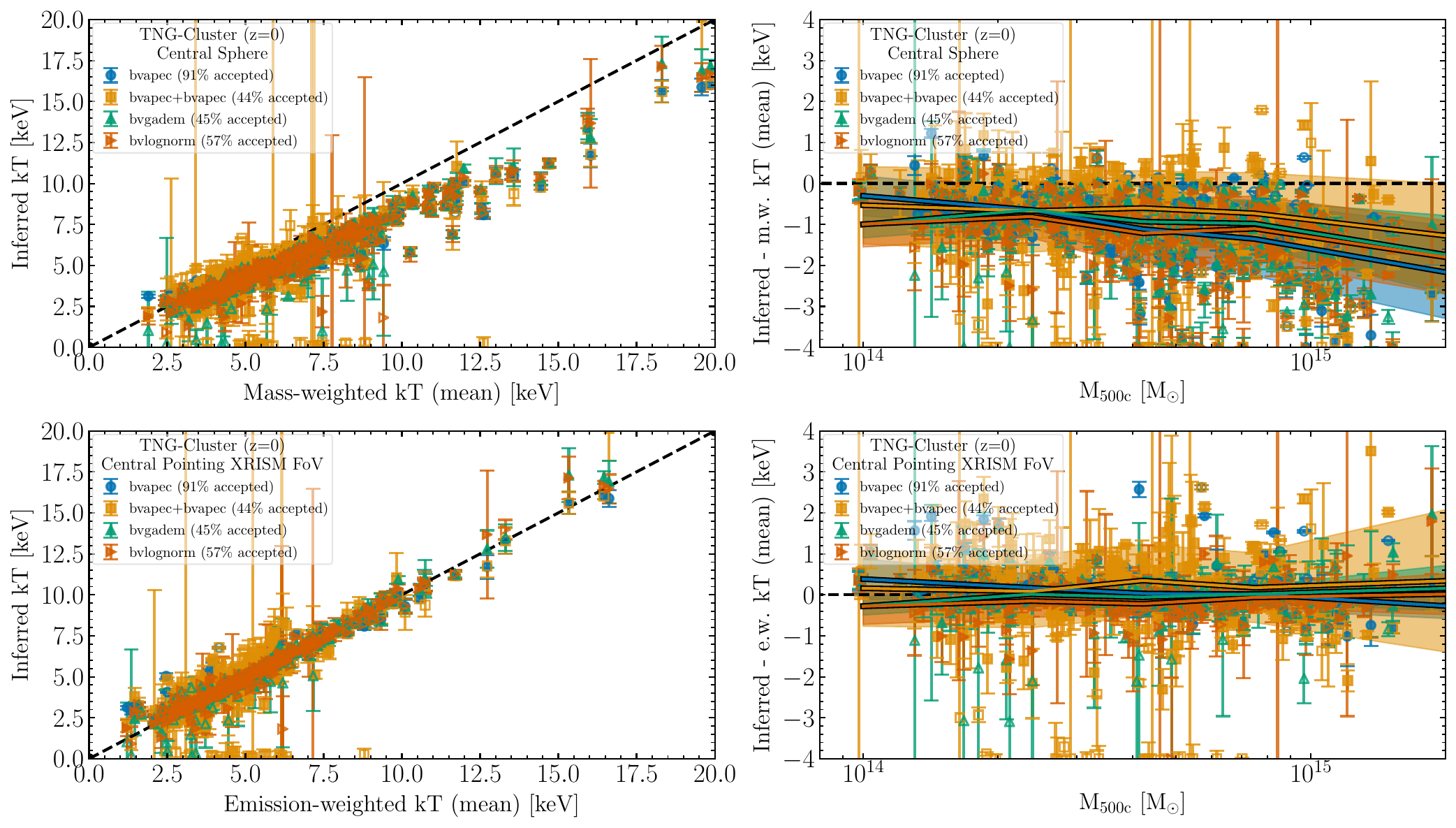}
    \caption{\textbf{Observational inferences of the ICM temperature at the centre of clusters.} In the left panels, we compare the temperature inferred by analysing the X-ray spectra from \textit{XRISM}/Resolve-like observations by adopting a variety of spectral-emission models (different colours) to the true mean temperature of the input temperature distribution, i.e. of systems simulated in TNG-Cluster. In particular, as ground truth, we first consider the mass-weighted gas in the 3D clusters’ centres (top) and then the emission-weighted gas for a central pointing with a \textit{XRISM}-like FoV (bottom). One marker represents one simulated cluster, in a random projection. Filled and empty symbols represent accepted and rejected best fits, respectively (as defined in Section~\ref{sec:accept}). In the right panels, we indicate the inference offset (or inference error) as a function of each cluster's M$_\text{500c}$ halo mass. Solid lines and shaded regions represent the running average and 1$\sigma$ standard deviation of the difference in bins of \text{log}$_{10}$(\text{M$_{500\text{c}}$}/M$_\odot$)=0.25, accounting only for the accepted fits. In all panels, the black dashed line represents unity, i.e. perfect accuracy of the model inference. Our results indicate that all models, are capable of correctly inferring the emission-weighted mean temperature of the gas in the targeted FoV, i.e. accounting for all gas on the line of sight. However, no model returns a best-fit ICM temperature value that is representative of the mass-weighted temperature of the central ICM gas to better than $1-2$ keV on average.}
    \label{fig:Temp_inf}
\end{figure*}

\begin{table}
    \centering
    \begin{tabular}{l c c}
    \hline \hline
    Model & Accepted fits & Acceptance rate (per cent) \\
    \hline
    \texttt{bvapec} & 322 & 91\\
    double \texttt{bvapec} & 154 & 44\\
    \texttt{bvgadem} & 160 & 45\\
    \texttt{bvlognorm} & 199 & 57\\
    \hline \hline
    \end{tabular}
    \caption{\textbf{Performance rates of each spectral-emission model adopted to analyse the X-ray spectrum of the central ICM in clusters simulated with TNG-Cluster.} We report the number of clusters, among a total of 352 at $z=0$, for which each model is able to produce an accepted fit for a 100~ks \textit{XRISM}/Resolve-like observation, following the criteria in Section~\ref{sec:accept}. We also indicate the percentage of valid fits compared to the total, hence indicating their relative acceptance rate. The models that assume unimodal temperature distributions with some width produce accepted fits more often than the double \texttt{bvapec} one.}
    \label{tab:rates}
\end{table}

As shown and discussed above, as well as in \cite{Rohr24b} and \citet{Staffehl25}, the intrinsic temperature distribution of the ICM, as predicted by TNG-Cluster, is multiphase in nature. Therefore, an important aspect for the accuracy of spectral-emission models is their ability to reflect such a multiphaseness in their best fits -- we tackle this next. 

First, before quantifying the inference errors, we comment on how frequently it is even possible to successfully fit the X-ray spectra of \textit{XRISM}/Resolve-like observations of the centres of clusters simulated in TNG-Cluster as a function of underlying spectral-modelling choices. Throughout, we adopt the mocking procedure described in Section~\ref{sec:mock_proc} and the fitting analysis and operational definitions of ``accepted'' and ``rejected'' fits of Section~\ref{sec:mock_analysis}.

Based on the success rates of the various models reported in \autoref{tab:rates}, the \texttt{bvlognorm} model is the best performer among the multi-temperature ones, followed by the \texttt{bvgadem} model and the double \texttt{bvapec}, in that order. The \texttt{bvapec} model has, by far, the largest number of accepted fits: however, by assuming a delta function for the temperature distribution, it fails a priori to provide a realistic depiction of the ICM's thermal properties. For the other models, the success rate can be taken as a measure for the compatibility between the intrinsic and the assumed temperature distributions. For example, the \texttt{bvlognorm} model assumes a more accurate depiction of the ICM temperature distribution as a unimodal distribution and remains insensitive to the sub-keV gas, which should have a negligible contribution in a \textit{XRISM}/Resolve like spectrum. On the other hand, the \texttt{bvgadem} model, which tends to overestimate the contribution of the sub-keV gas, underperforms due to inflated distribution widths. With a similar acceptance rate, the double \texttt{bvapec} model also underperforms since the simulated clusters lack any visible bimodality, i.e. two similarly-prominent peaks.

\begin{figure}
    \centering
    \includegraphics[width=\columnwidth]{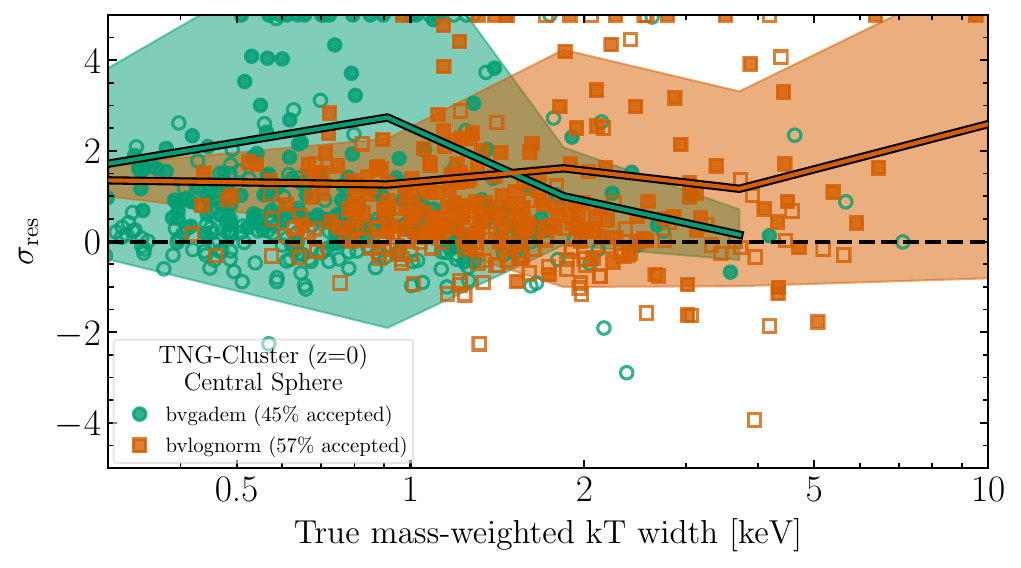}
    \includegraphics[width=\columnwidth]{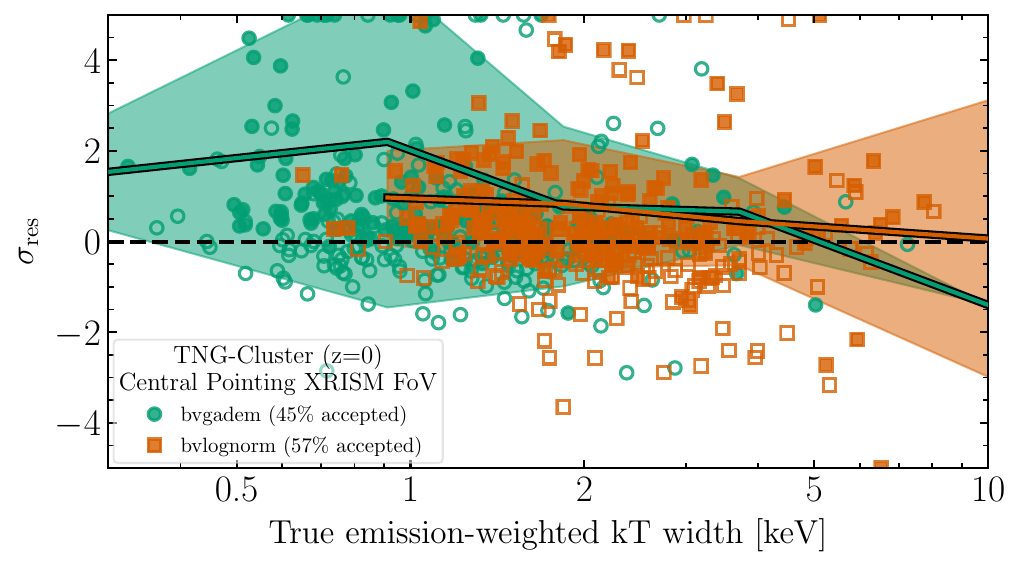}
    \caption{\textbf{Observational inferences of the width of the temperature distributions of the central ICM.} We quantify the residuals of the temperature width for the \texttt{bvgadem} and \texttt{bvlognorm} models (Equation~\ref{eq:sigma_res}) as a function of the ``true'' width of the temperature distributions for the mass-weighted central gas (top) or the emission-weighted gas for a central pointing with a \textit{XRISM}-like FoV (bottom), for the same X-ray spectra analysis of Figure~\ref{fig:Temp_inf}. Filled and empty symbols show accepted and rejected best fits per-cluster, respectively, while the black dashed line represents unity. Solid lines and shaded regions represent the running average and 1$\sigma$ standard deviation of the $\sigma_{\rm{res}}$ in bins of log$_{10}(M_{\rm{500c}}/M_\odot)$=0.25, accounting only for the accepted fits. Our results suggest that the \texttt{bvlognorm} model, compared to \texttt{bvgadem}, is systematically returning a more accurate description of the temperature distribution width, but both models struggle in capturing the true extent of the ICM's temperature distribution at the centre of clusters (top).}
    \label{fig:width_inf}
\end{figure}

Regarding accuracy, we assess first the best-fit values for the primary temperature value (or ICM effective temperature, \autoref{fig:Temp_inf}) and then, when appropriate, the width of the distribution (\autoref{fig:width_inf}). In both cases, we compare the best-fit parameters with the ``true'' values from the simulations by referring to both the mass-weighted gas properties of the ICM in a central sphere as well as the emission-weighted properties in a central pointing with \textit{XRISM}-like FoV.

{\it ICM effective temperature.} For the ``true'' ICM single-value temperature we take the mean of the temperature distribution from the simulated clusters, i.e. the mean of each input distribution. This is compared against the best-fit values. The mode or median could also be good summary statistics of the TNG-Cluster temperature distributions, but they do not represent the outputs of the \texttt{bvgadem} and \texttt{bvlognorm} models. For the double \texttt{bvapec} model, we consider exclusively the best-fit primary temperature, namely the one that maximizes the relative contribution to the spectrum. We note that for a handful of systems, the double \texttt{bvapec} model finds a hotter, instead of a colder, secondary i.e. subdominant component temperature.

Based on \autoref{fig:Temp_inf}, top panels, all models tend to under-predict the mass-weighted mean temperature of the X-ray emitting gas that is actually located in the centre of each cluster. Whereas the bias is systematic, it is in fact relatively small. All models, on average i.e. across the TNG-Cluster sample, underestimate the temperature by 1.19~keV (an error of $\sim$ 17 per cent), but with a strong dependence on cluster mass. Namely, the best-fit temperature can underestimate the true one by about 1.5~keV ($\sim$20 per cent) at the high-mass end, while being a reasonable approximation for lower-mass clusters. This result underscores the difficulty of assessing the true physical state of central ICM gas. In turn, assuming that TNG-Cluster returns reasonably realistic systems, the top panels of \autoref{fig:Temp_inf} can provide guidelines to correct for this observational-analysis bias. This is the case on average, whereas for individual clusters the inference error can be much larger -- we expand on outlier cases later.

In terms of emission-weighted quantities and comparing to the true properties of all the gas along the line of sight of a \textit{XRISM}-like central pointing (\autoref{fig:Temp_inf}, bottom panels), the situation improves. All models produce a best-fit temperature that is indicative of the emission-weighted mean temperature of the gas for a large majority of the studied clusters, independent of halo mass. For the double \texttt{bvapec} model, in particular, and for both the top and bottom panels, the inference errors span the largest ranges, with a sizeable fraction of clusters whose observational-based (primary) temperature is substantially higher or lower than the true one. Moreover, the errors on the best-fit values for the double \texttt{bvapec} model are on average much larger than in any other case.

We find no significant differences in the quality of the inferences between the accepted and rejected fits, at least according to our definitions. In all cases, they occupy the same parameter space, suggesting that the rejection of the fit does not depend on the inference of the mean mass- or emission-weighted temperature.

{\it ICM temperature width.} Beyond the mean temperature, a key parameter in the description of the ICM temperature structure is the width of the distribution. As the ``true'' value we take the standard deviations of the best-fit Normal and Log-Normal distributions to the intrinsic gas (\autoref{fig:gas_prop_fit} and \autoref{fig:best_fit_val}). We distinguish between the results for the mass-weighted gas in the 3D cluster centre versus emission-weighted gas for a central pointing with a \textit{XRISM}-like FoV. We then compare these values to the inferred widths from the spectral fitting assuming, respectively, the \texttt{bvgadem} and \texttt{bvlognorm} spectral-emission models. 

In order to assess both the accuracy and the significance of the observational inferences on the underlying thermal properties of the gas, in \autoref{fig:width_inf} we show the residuals of the inferred thermal width as a function of the ground truth:
\begin{equation}
    \sigma_\text{{res,i}}=\frac{\sigma_\text{inf,i}-\sigma_\text{true,i}}{\Delta\sigma_\text{inf,i}}, ~~\text{i}=\text{Normal~or~Log-Normal}
    \label{eq:sigma_res}
\end{equation}
where $\sigma_\text{inf}$ and $\Delta\sigma_\text{inf}$ refer to the distribution width and its 1$\sigma$ error as inferred by fitting the spectra and where $\sigma_\text{true}$ is the true width of the temperature distribution, as predicted by TNG-Cluster. These residuals so defined provide a gauge for the accuracy of the inferred temperature distribution width, at least for the accepted fits; on the other hand, the values in the rejected cases are primarily driven by the denominator and are representative of the higher uncertainties associated with a spectral fit that has not converged.

In \autoref{fig:width_inf}, we quantify the residuals for both the mass-weighted temperature distribution of the central gas (top) and the emission-weighted gas in a \textit{XRISM}-like FoV (bottom). For both cases, the \texttt{bvgadem} model tends to overestimate the width of the distribution, with an average $\sigma_{\rm{res}}$ of 1.56 (-1.77,+0.55) and 1.27 (-1.67,+0.56), for top and bottom cases, respectively. On the other hand, the \texttt{bvlognorm} model tends to overestimate the average $\sigma_{\rm{res}}$ for the central mass-weighted gas (1.05 (-1.28,+0.70)), but is in a good agreement with the width of the emission-weighted gas in the \textit{XRISM} FoV by an average of 0.41 (-1.05,+0.85). We note that, for both models, we find inferences that deviate by more than 5$\sigma$ from the true width of the distribution, even for accepted fits. These cases appear to be systems with temperature distributions that have pronounced tails of cooler or super-virial gas.

Overall, the \texttt{bvlognorm} spectral-emission model can better capture the true width of the broad temperature distribution of the central ICM gas, whether mass-weighted or emission-weighted and including projected contributions.

\begin{figure*}
    \centering
    \includegraphics[width=\linewidth]{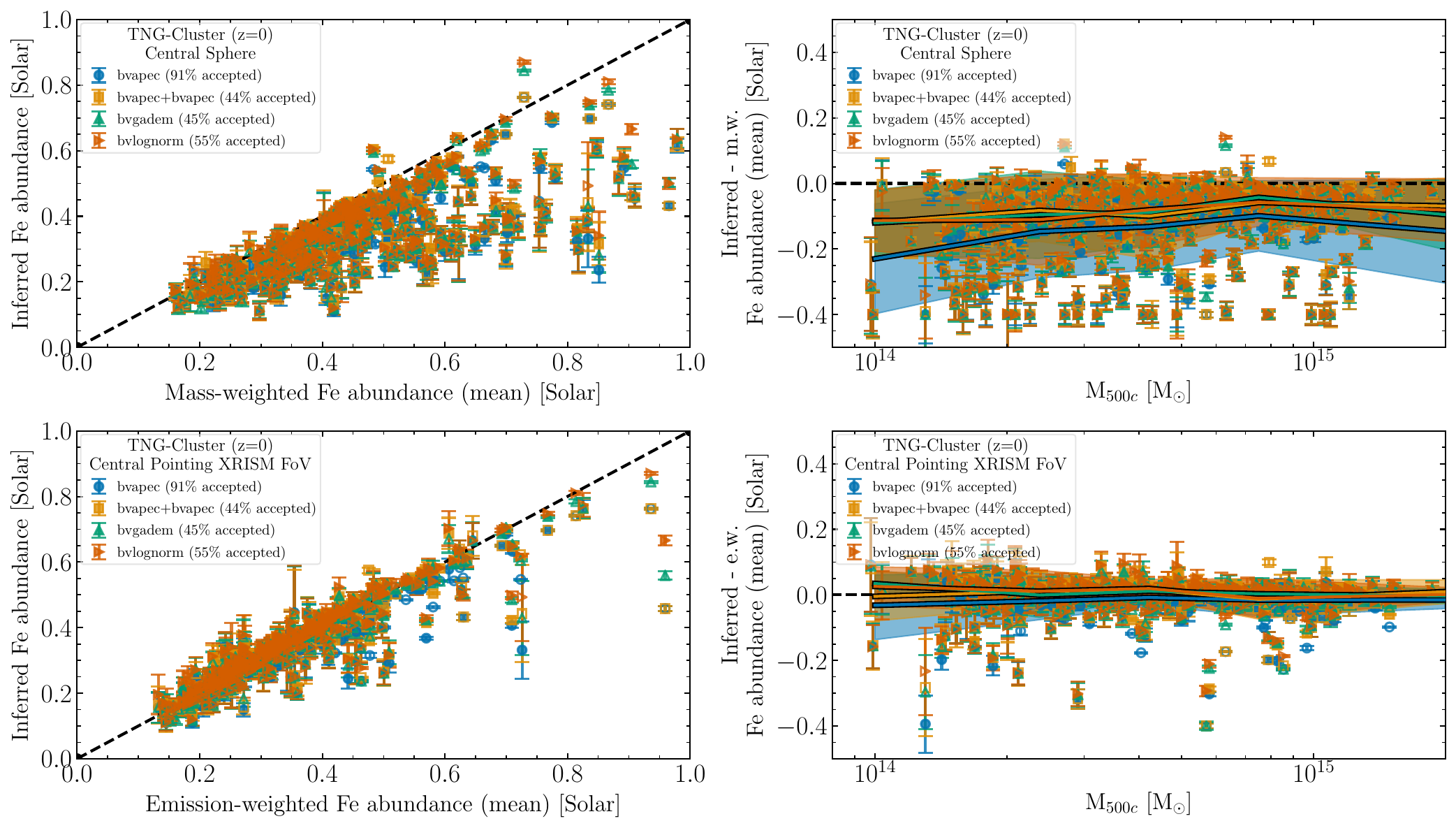}
    \caption{\textbf{Observational inferences of the Fe abundance of the ICM at the centre of clusters.} For the same spectral fits and modelling of \autoref{fig:Temp_inf}, we compare the Fe abundance inferred by all spectral-emission models used in this work to the true mean Fe abundance (left panels) of the input distribution: mass-weighted central gas (top) and  emission-weighted gas for a central pointing with a \textit{XRISM}-like FoV (bottom). Annotations are as in \autoref{fig:Temp_inf} and, as there, we indicate the inference offset as a function of each cluster's M$_\text{500c}$ halo mass (right panels). Solid lines and shaded regions represent the running average and 1$\sigma$ standard deviation of the difference in bins of log$_{10}(M_{\rm{500c}}/M_\odot)$=0.25, accounting only for the accepted fits. Our results indicate that all models correctly infer the mean, emission-weighted Fe abundance of the gas in the FoV. However, all inferences systematically underestimate the average true amount of Fe in the clusters' centres.}
    \label{fig:Fe_inf}
\end{figure*}

\subsection{Inferring the ICM Fe abundance}\label{sec:Fe_abund}

We proceed to assess how well observational inferences of the ICM Fe abundance return the ``true'' underlying values, by following the same procedure as for the temperature, in fact by inspecting the results of the same fits of \textit{XRISM}/Resolve-like spectra from TNG-Cluster discussed thus far. 

In \autoref{fig:Fe_inf} we show the comparison between the inferred Fe abundance from all spectral-emission models compared to the mass-weighted mean Fe abundance of the ICM in the actual 3D cluster centre (top) and the emission-weighted mean Fe abundance of gas from a central pointing with a \textit{XRISM}-like FoV (bottom). As the top panel of \autoref{fig:Fe_inf} shows, the inference on the Fe abundance is systematically different of the actual amount of Fe in cluster centres. Namely, all models underestimate the central Fe abundance by about 0.12~Solar (22 per cent). The offset varies with halo mass, being larger towards the low-mass end of our mass range. It is important to point out that such a systematic offset is significantly larger than the statistical errors on the Fe abundance typically quoted in the literature, of the order of 0.01, e.g. for the inferences from \textit{XRISM}/Resolve observations available in the literature.

However, the inference error is at least partially driven by the subset of clusters that, despite being actually metal rich, are found to be significantly metal poorer when their X-ray spectra are analysed, irrespective of the adopted spectral-emission model. This discrepancy disappears if we account for emission (bottom panels), suggesting it is possibly driven by less X-ray bright gas. Interestingly, even though the rejected fits do not occupy a special region in the parameter space, as in the temperature case, a disproportional amount of clusters appear to belong to this subset.

Similar to the case of the emission-weighted mean temperature, all models consistently trace the emission-weighted Fe abundance, or at least they trace more accurately the emission-weighted than the mass-weighted Fe abundance, with no significant halo-mass dependence.

\begin{figure*}
\centering
        \includegraphics[width=0.9\textwidth]{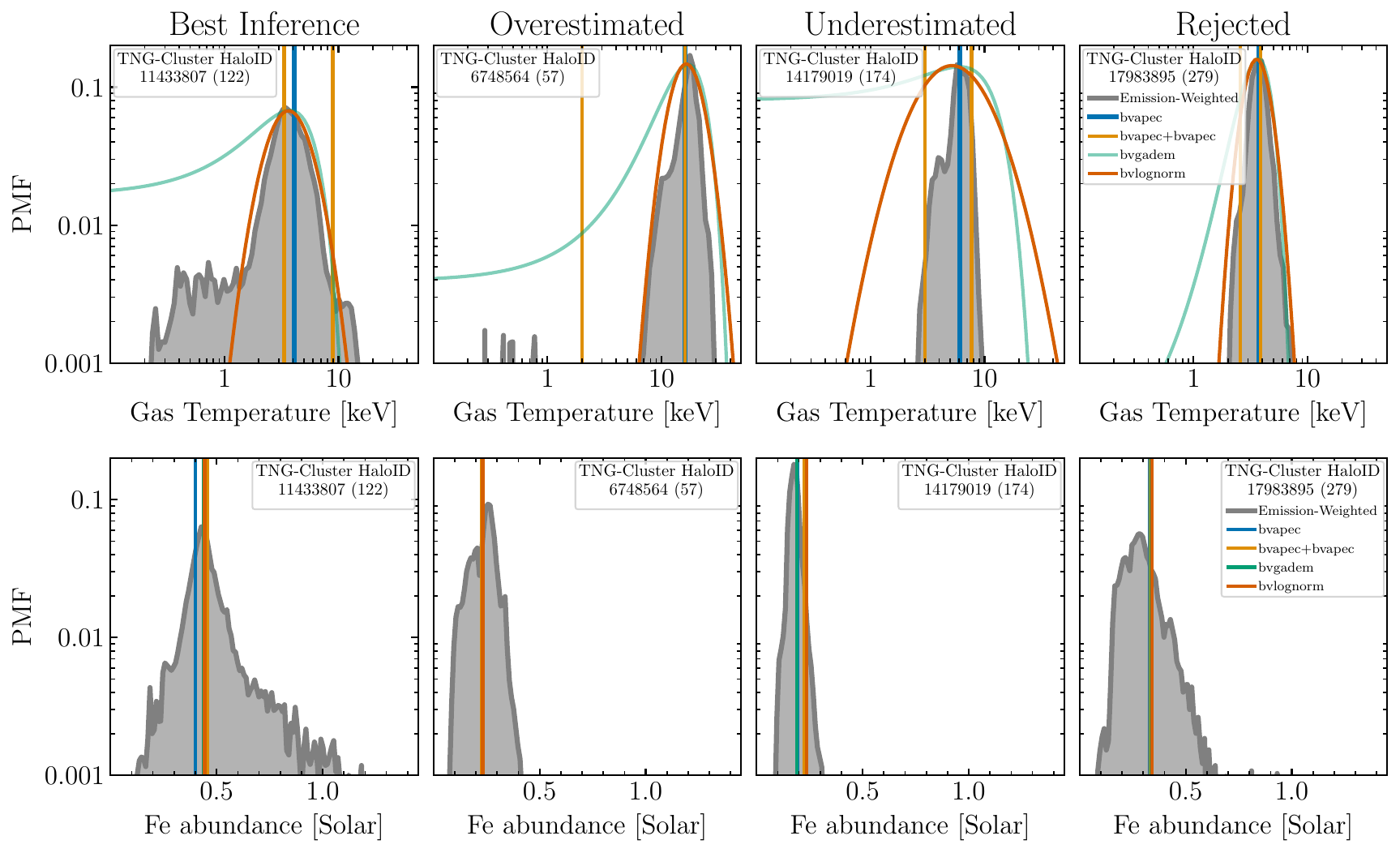}
        \includegraphics[width=0.9\textwidth]{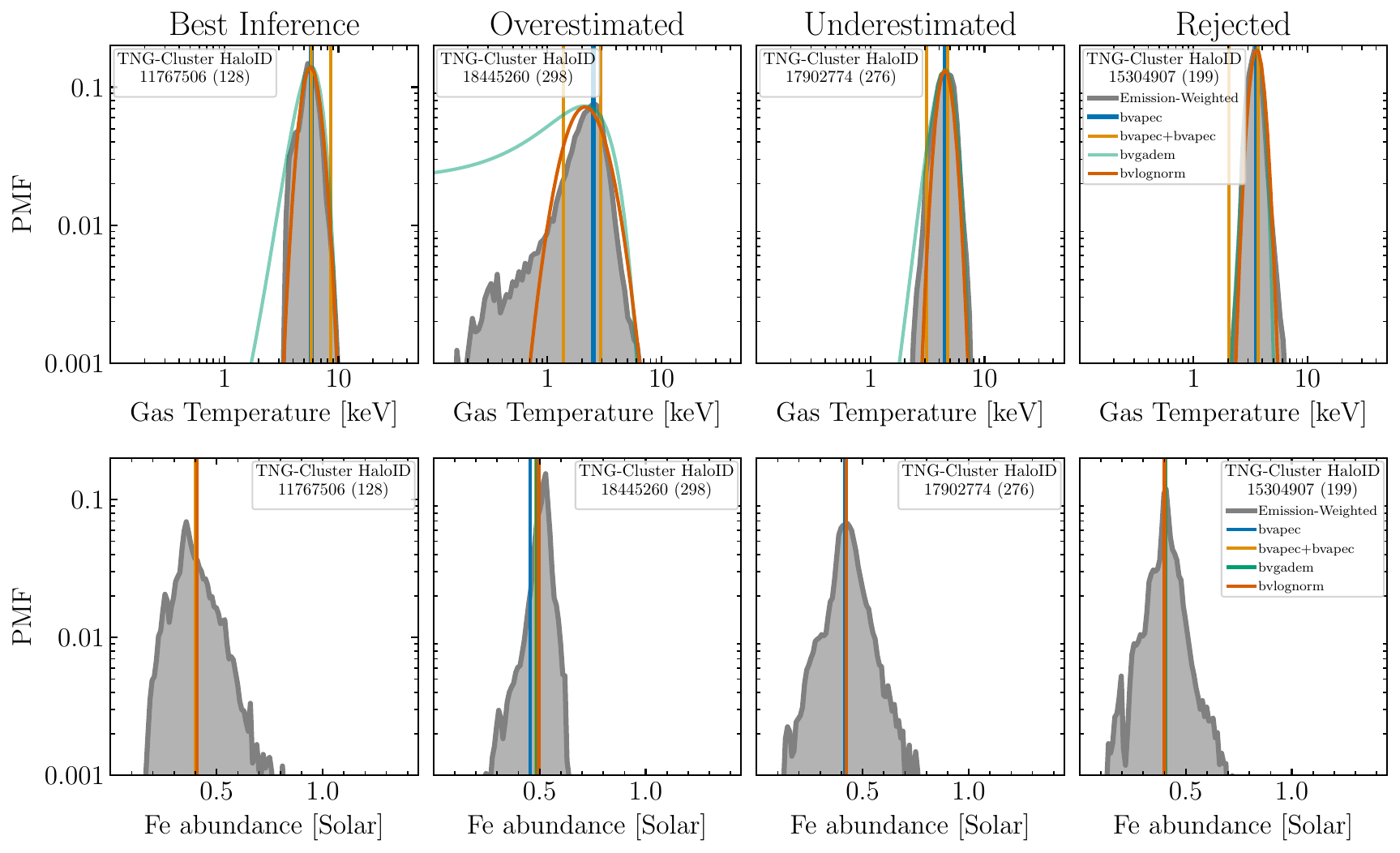}
    \caption{\textbf{Observational inferences compared to the intrinsic gas cell distributions for a few TNG-Cluster examples.} We present the true emission-weighted distributions of temperature and Fe abundance of the gas cells, as a fraction of the EM per bin, for a central pointing with a \textit{XRISM}-like FoV from TNG-Cluster in grey as they compared to the outcome of X-ray fitting via each of the spectral-emission models studied in this work. The individual example clusters shown in this Figure are selected based on the quality of the \texttt{bvlognorm} inference w.r.t. the mean of the temperature (top), mean Fe abundance (bottom; continued) and the best-fit Log-Normal distribution width (middle). For each case we showcase a system with a best, an overestimated and underestimated inference, from left to right. A case of a rejected fit with a reasonable inference is also shown as a comparison (rightmost column). Our results highlight the strength and weaknesses of each model in providing an accurate representation of the complex input distributions of the gas in clusters (see text for details). This Figure also provides examples whereby the lack of constraints on the breadth of the ICM temperature distribution as the main factor behind rejection of both \texttt{bvgadem} and \texttt{bvlognorm} models.}
    \label{fig:inf}
\end{figure*}

\begin{figure*}
\ContinuedFloat\centering
        \includegraphics[width=0.9\textwidth]{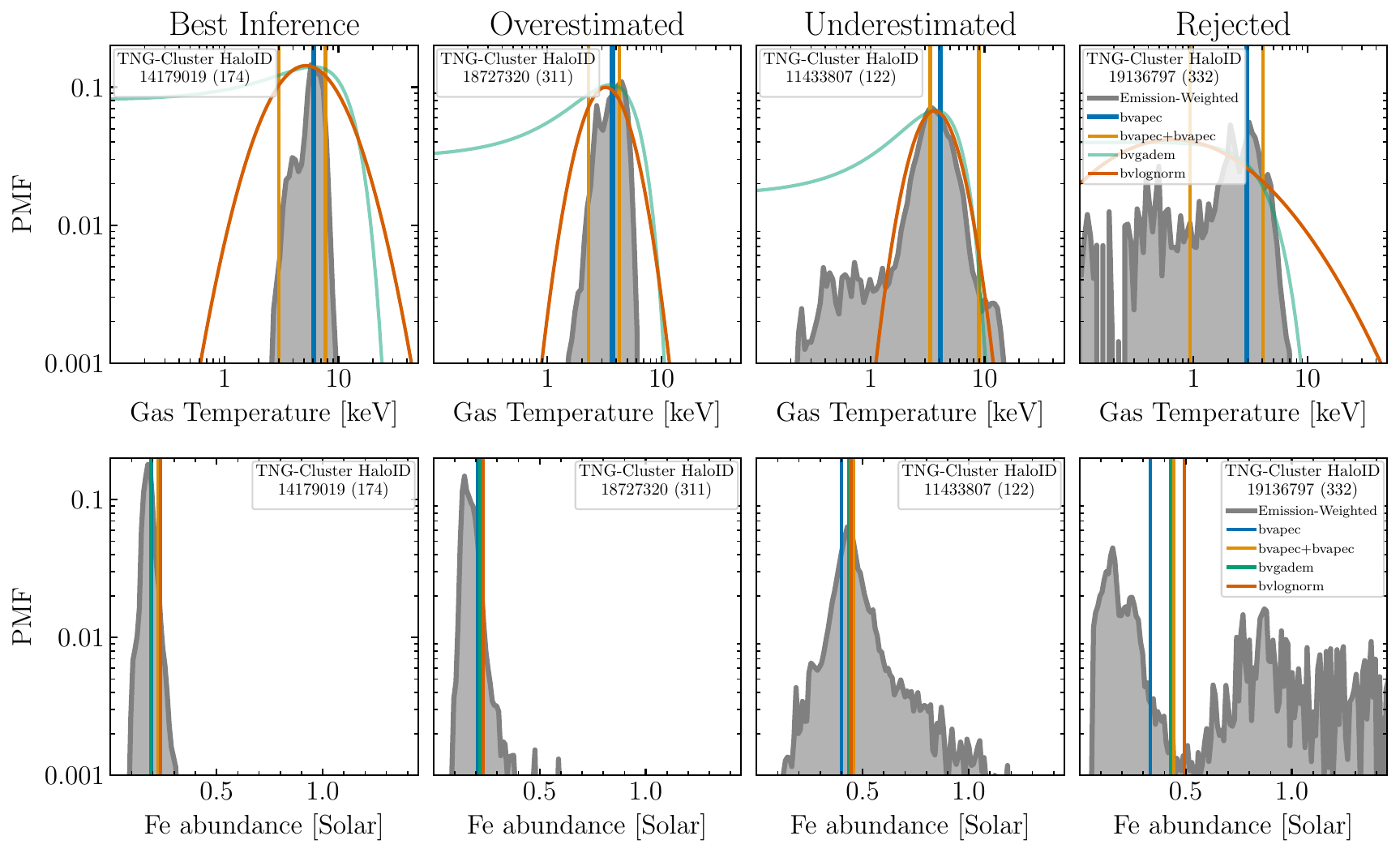}
        \caption{Continuation of \autoref{fig:inf}}
\end{figure*}

Overall, all models tend to underestimate the average Fe abundance of the gas in the centre of clusters. However, the single-point estimate for the amount of Fe is too simplistic of an approximation of the true, extended distribution. We speculate, for example, that extrapolating this value throughout the entire volume could still systematically inflate the total amount of Fe in the ICM.


\section{Discussion}\label{sec:discussion}

Overall, our results indicate that the assumptions on the temperature distribution of the gas may have an important impact on the inference of the properties of the ICM at the centre of clusters from X-ray spectral fitting, at least of those predicted by TNG-Cluster. 

As shown in \autoref{fig:Temp_inf} and \autoref{fig:Fe_inf}, the inferences of the \texttt{bvapec} model on the emission-weighted temperature and Fe abundance of the gas along the LoS are comparable to (if not more accurate on average than) the other spectral-emission models. We note that for the temperature, this is achieved despite us not using the so-called spectroscopic temperature weighting from \citet{Mazzotta04}, which was introduced to account for biases on the inference from CCD-resolution mock observations. Similarly, we find a good agreement between the true and inferred Fe abundance, even for haloes with temperatures between 2 and 3~keV, suggesting that our estimates are not subjected to the "inverse Fe-bias" \citep[e.g.][]{Rasia08,Simionescu09,Gastaldello10}. We believe that, in both cases, the \texttt{bvapec} model is able to provide a more accurate inference on the true temperature and Fe abundance of the ICM due to the higher spectral resolution, which reveals more lines that can place strong constraints on the model. However, our analysis remains limited to the cluster centres and the quality of the inference varies from target to target, suggesting that a broader examination of this topic, for a large sample of galaxy clusters over larger scales is necessary, but outside the scope of this project.

However, the \texttt{bvapec} model, by design, ignores the presence of multi-phase gas and cannot provide us any insights on the distribution of temperatures present in the ICM. Attempting to rectify this with the inclusion of another single-temperature component, in the form of the double \texttt{bvapec} model, does not improve the accuracy of the inferences on average (i.e. across the studied cluster sample) and actually returns a much larger fraction of clusters with very large inference errors on the ICM effective temperature, with biases in both directions.

The \texttt{bvgadem} and \texttt{bvlognorm} assume a Normal and Log-Normal distribution of temperatures, respectively, hence allowing a more complete description of the properties and diversity of the gas at the centre of clusters. For this reason, they return somewhat more frequently successful fits than the double \texttt{bvapec} model. And for the same reason, we argue that they should be prioritized, especially when attempting to study complex phenomena such as the effects of AGN feedback on the ICM, which could impact different gas phases differently \citep[as is the case in the TNG model, see e.g. ][]{Truong24}. However, on average, the \texttt{bvgadem} and \texttt{bvlognorm} models do not necessarily return more accurate inferences of the ICM effective temperature and average Fe abundance. And, as shown, they too slightly albeit systematically underestimate the bulk properties of the ICM that is actually located close to the cluster centre.

In the following sections, we compare the fits produced by each model to the simulation-intrinsic distributions, in order to examine the quality of each inference in greater detail. We also investigate the impact that projection effects and brightness bias have on the inference of the properties of the central gas, in order to understand the systematic offsets between the two quantified so far. Finally, we also attempt to account for selection biases, such as focusing only on the X-ray brighter strong cool-core (SCC) clusters, in order to make our results more applicable to typical observational approaches.

\begin{figure*}
\centering
        \includegraphics[height=0.45\textwidth]{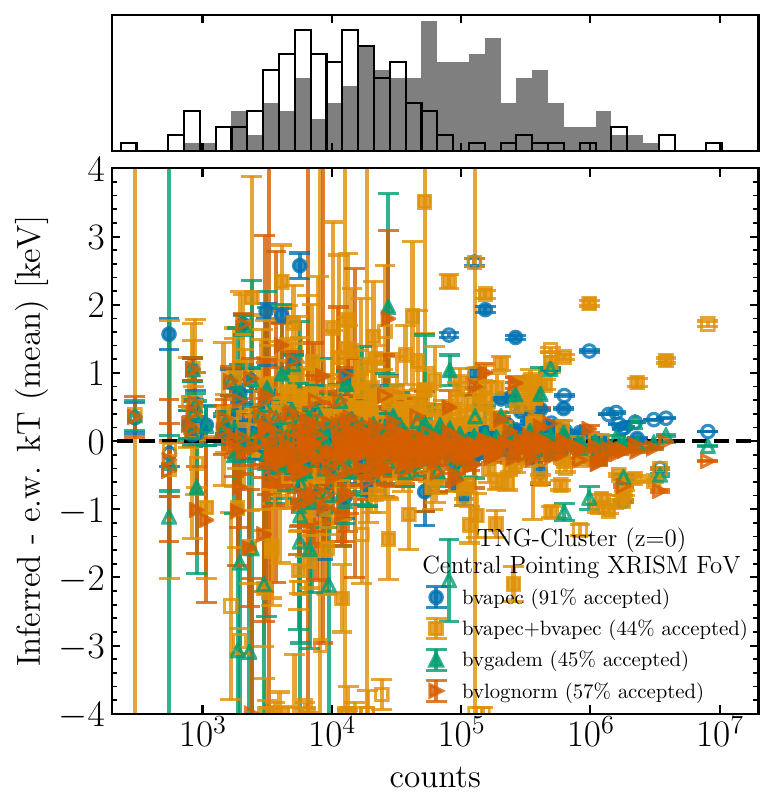}
        \includegraphics[height=0.45\textwidth]{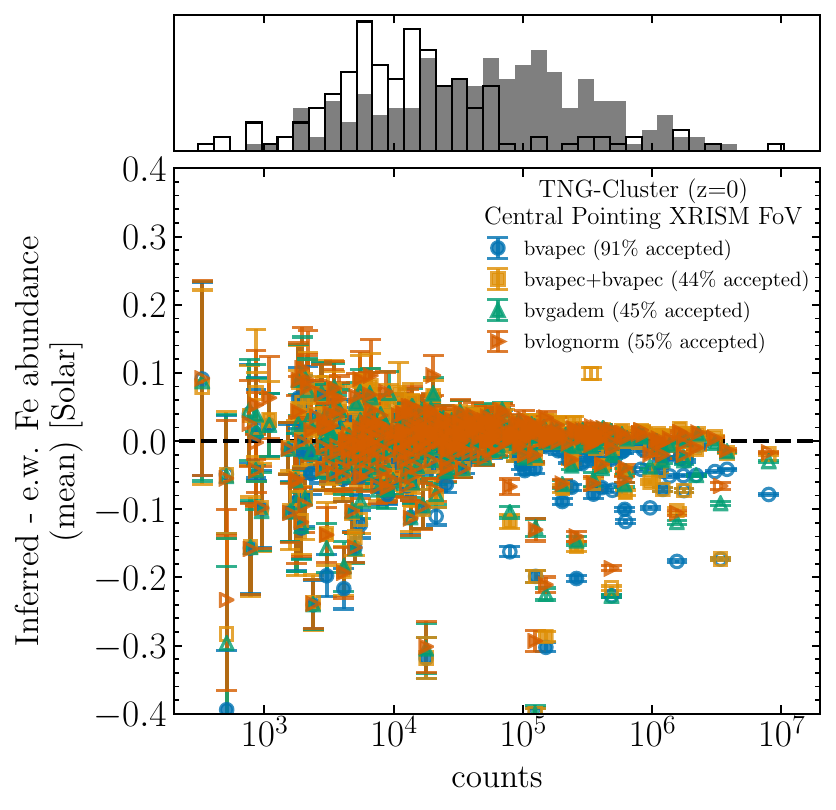}
    \caption{\textbf{Inferring the ICM temperature and Fe abundance, across spectral-emission models, as a function of detected photons per cluster.} We present the quality of each model's inference on the emission-weighted mean temperature (left) and Fe abundance (right) of the gas as a function of the number of counts observed for a 100~ks exposure. As previously, we consider all the ICM for a central pointing with a \textit{XRISM}-like FoV. The black dashed line represents unity. We present the accepted and rejected fits with filled and open symbols, respectively. In the two top subplots, we also show the distribution of photon counts per clusters for the accepted (filled) and rejected (empty) \texttt{bvlognorm} best fits. Our findings indicate that the best fits of fainter clusters tend to be more likely to be rejected, suggesting a correlation between acceptance and X-ray brightness, and even if rejected and accepted fits seem to return similarly-accurate results across the cluster sample}
    \label{fig:counts}
\end{figure*}

\subsection{Insights on the quality of the observational inferences, and of the fits, across spectral-emission models}\label{sec:insights}

As we have previously shown, all the spectral-emission models used in this work are capable of inferring with good accuracy the emission-weighted summary statistics of the cluster gas from a central pointing with a \textit{XRISM}-like observation. Yet, for a large number of clusters, they fail to return an accepted best fit of the underlying X-ray spectrum. Moreover, as mentioned above and by comparing the filled and empty markers of \autoref{fig:Temp_inf}, \autoref{fig:width_inf} and \autoref{fig:Fe_inf}, it would appear that the rejection of the X-ray spectra fits (at least based on our criteria, see Section~\ref{sec:accept}) does not translate in much or overall worse inference errors. This can be appreciated via the quantitative summaries provided in \autoref{tab:errors_1} and \autoref{tab:errors_2}. In the following, we hence provide further considerations as to why the quality of the fit may still be relevant when choosing which spectral-emission model to adopt to analyse the data.

\subsubsection{Examples of true vs. inferred gas distributions}
In \autoref{fig:inf} we showcase the {\it emission-weighted} distributions of the ICM properties of a few example individual systems in comparison to their best-fit description, for all the spectral-emission models studied in this work. Each cluster is represented by two panels, one for temperature and one for Fe abundance, respectively, on top of each other. We show a total of 12 example clusters, selected based on the performance of the \texttt{bvlognorm} model to infer, respectively, effective temperature (top two rows), Fe abundance (bottom two rows; continued Figure) and temperature width (middle two rows). 

We focus on the \texttt{bvlognorm} because it provides a good descriptor of the TNG-Cluster temperature distributions  (\autoref{sec:intrinsic}) and because it provides more frequently more accurate best-fit results for the temperature distribution widths (\autoref{sec:temp}). However, this discussion can be generalized for all spectral-emission models.

From left to right, we visualize the intrinsic and inferred gas property distributions for different example cases. Similar to \autoref{fig:gas_prop_fit}, we present the probability mass function of the emission-weighted gas distribution along the LoS as a fraction of the EM per bin. In the leftmost column, we show systems with the minimum absolute difference between the true and inferred value of the relevant property. We also showcase instances of an overestimation and underestimation, located at the 95$^{\text{th}}$ and 5$^{\text{th}}$ percentile of the sorted difference distribution, respectively (middle columns). Finally, we show examples of rejected fits (rightmost column), so that we can better see the intrinsic properties of the gas that, despite an accurate inference, led to model rejection. 

Starting with the accepted spectral fits, we can understand how each model attempts to infer the input temperature distribution. The double \texttt{bvapec} model systematically infers temperatures that are not present or not associated with any minor peaks in the temperature distribution. Similarly, in many cases it fails to identify the most prominent peak, by inferring two temperatures around it. This suggests that the model, in an attempt to approximate a bimodal distribution of temperatures, does not provide an accurate representation of the actual distribution, but an attempt of residual minimisation. 

The \texttt{bvgadem} model is capable of determining the peak of the temperature distribution, but systematically assumes a much larger amount of cooler gas compared to what is present. Finally, many cases show how the \texttt{bvlognorm} model provides a fairly accurate representation of the width of the temperature distributions. However, it tends to fail when the intrinsic temperature distribution deviates strongly from a Log-Normal distribution in the form of extended tails towards higher- or lower-temperature gas.

Regarding the Fe abundance distributions, all spectral-emission models make overly simplistic approximations. We see a range in the capability of all models to identify the mean Fe abundance, consistent with the trends found above. 

Another important aspect regarding the inference of each model are the conditions that led to its rejection. As we can see in \autoref{fig:inf} and as previously indicated in \autoref{fig:Temp_inf} and \autoref{fig:Fe_inf}, even rejected fits of all models are capable of inferring a fairly accurate value of the mean temperature and Fe abundance emission-weighted distribution. For \texttt{bvgadem} and \texttt{bvlognorm}, the main and most frequent factor that leads to the rejection of the model is the width of the distribution: in some rejected cases, the inferred width is simply too large, infinity; in other rejected cases, the model infers a very narrow temperature distribution, i.e. with a poor constraint on the width itself.

\subsubsection{Dependence of quality of the fit on number of photons}
Considering the sensitivity of all models on the emission-weighted properties of the gas along the LoS, it is possible that a model cannot infer the temperatures of the less luminous gas contributing to the distribution because of a lack of counts. 

In \autoref{fig:counts} we examine this hypothesis by comparing each model inference on the temperature and Fe abundance as a function of the number of counts. It is evident from the distribution of our clusters that the rejected \texttt{bvlognorm} best-fits tend to have systematically lower counts than the accepted ones, but the count distributions of the two populations have a significant overlap. We note that the same behaviour holds for the accepted and rejected best-fits of the other spectral-emission models.

To understand the extent that exposure time may have on the rejection of our best-fits, we repeat the X-ray spectra mocking and analysis for the rejected clusters by increasing the exposure time to 300~ks, instead of the fiducial 100~ks adopted throughout. For the case of the \texttt{bvlognrom} model (which returns the largest success rate), we find that 55 per cent of the clusters that have been previously rejected turn out to have a valid fit with an increased exposure (not shown). This confirms the hypothesis that the rejection of a spectral fit for the fainter clusters in our sample is due to available exposure time.

However, this is not just a cluster-mass effect, as in fact there still remains a population of systems for which the best-fit is still rejected, even though their inferences on the emission-weighted mean temperature and Fe abundance are comparable with the rest of the sample. We find (although do not show) that clusters that are rejected even for a three-time longer exposure are either metal poor, with the majority of the emission-weighted mass in the ICM having an Fe abundance of $\sim$0.25~Solar, or deviate heavily from a Log-Normal distribution, with a significant amount of sub-keV gas. We note that such cases appear as almost separate populations, without any apparent correlation with halo mass (\autoref{fig:Fe_inf}, bottom left panel).

Regarding the systems that deviate from the Log-Normal distribution, it appears that they get rejected due to their poor goodness-to-fit. It is evident that the spectral-emission model's parametrisation is not able to fully capture their temperature distribution, leading to poor fits, despite the increased number of counts. On the other hand, the rejection of the metal poor population is due to the fact that the spectral-emission models use the emission lines also to recover the temperature distribution of the gas. Whereas the shape of the continuum is a reliable diagnostic of the ICM's bulk temperature, for plasma with temperatures higher than 2~keV, the emission-line ratios trace the presence of cooler or hotter gas, constraining the width of the distribution. However, if clusters have low elemental abundances, they will return low numbers of counts in spectral lines irrespective of the longer exposures, and in turn the modelling will be unable to provide any meaningful constraints on the width of the temperature distribution, leading to their rejection.

Overall this indicates that even the more accurate assumption of a temperature distribution is not a sufficient representation of the properties of the ICM, as predicted by TNG-Cluster, both for its thermal and more importantly its chemical structure. This highlights the limitations of current spectral-emission models and the need for more sophisticated descriptions of the intrinsic properties of the gas in order to better understand the ICM.

\subsection{Effects of observational realism}\label{sec:obs_bias}

Even though it might be tempting to assume that the inferred temperature and Fe abundance from a central-pointing spectrum are representative of the local conditions of the ICM, our analysis suggests otherwise. As shown e.g. in \autoref{fig:Temp_inf} and \autoref{fig:Fe_inf}, we systematically find large inference errors on the mass-weighted state of the central ICM, but much smaller deviations from the emission-weighted properties of the projected gas along our LoS. Whereas this differentiated fit performance may not be surprising, it may have implications when searching for a physical interpretations of the data.

We now explore the individual sources of those differences and quantify their impact and systematic uncertainties. In the following sections we therefore attempt to understand and quantify the systematic uncertainties introduced by projection effects and the emission bias, by isolating the two.For the purposes of this analysis we focus again exclusively on the  \texttt{bvlognorm} model as a reference point. However, the qualitative effects that we find are consistent across all models.

\begin{figure*}
    \centering
    \includegraphics[width=\linewidth]{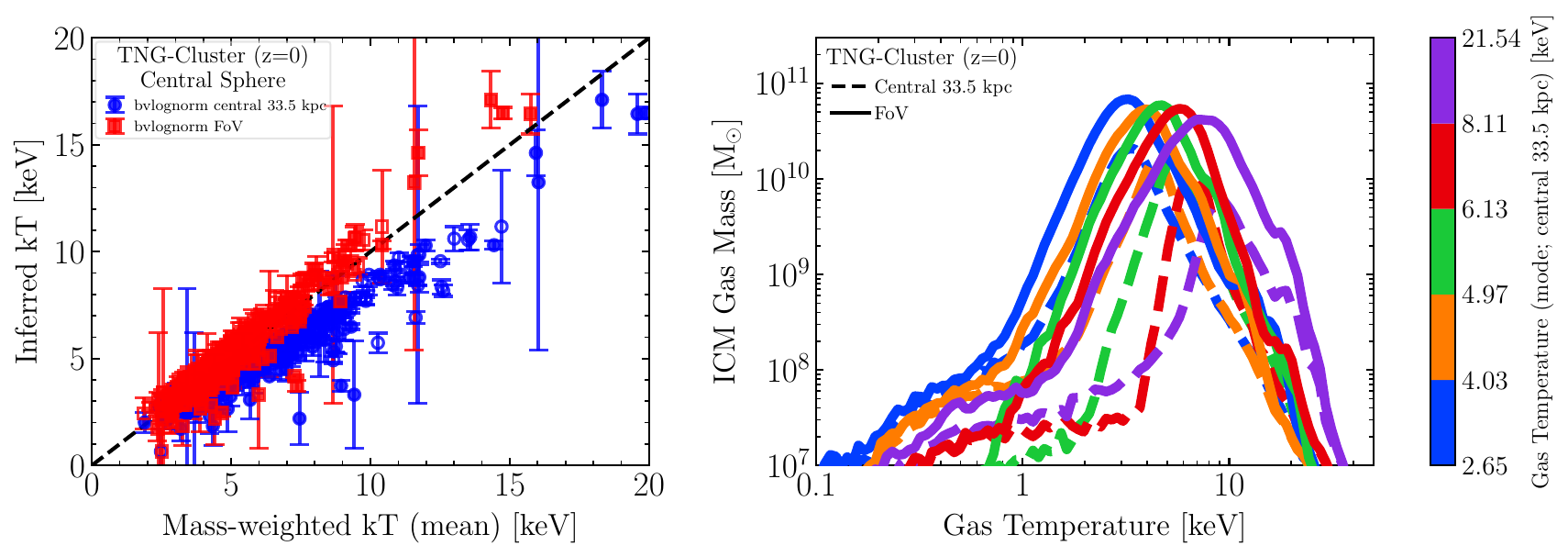}
    \includegraphics[width=\linewidth]{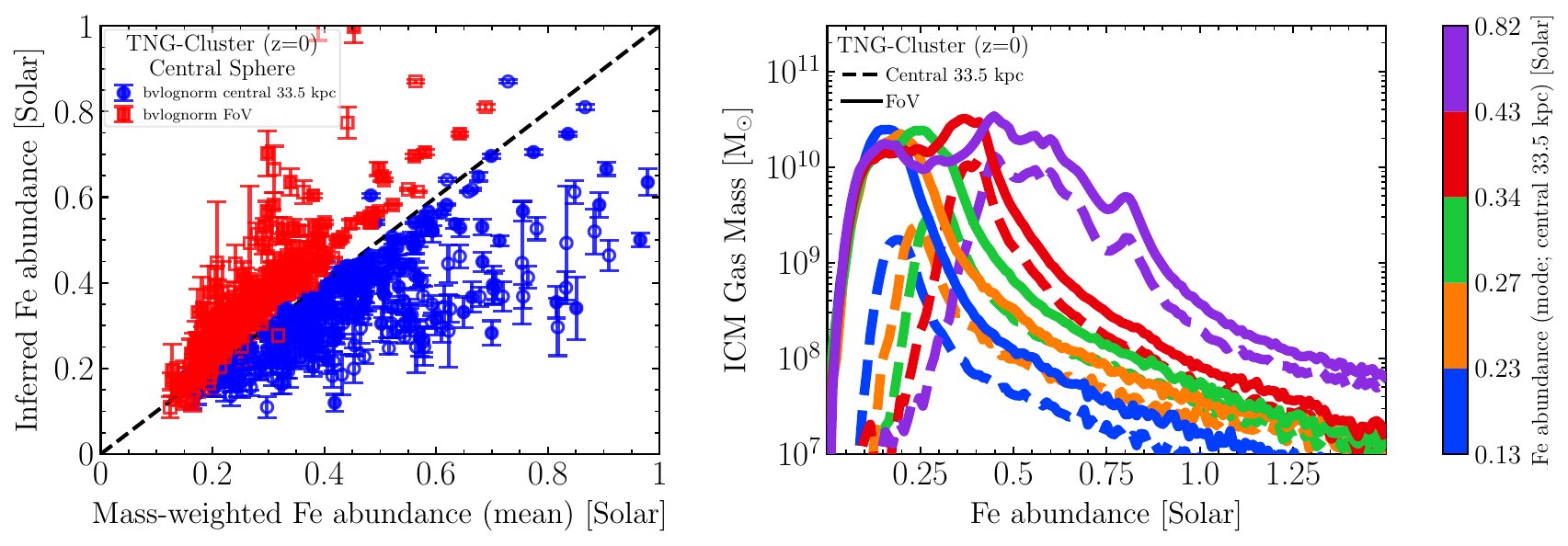}
    \caption{\textbf{Systematics due to projection effects in the temperature (top) and Fe abundance (bottom) distributions.} We highlight the impact of projection effects on the observational inferences, by comparing the \texttt{bvlognorm} best-fit outcome to the true mass-weighted mean temperature of the central (blue) vs. projected (red) gas (left). The latter includes all gas for a central pointing with a \textit{XRISM}-like FoV but, differently than in previous Figures, without weighting by X-ray emission. The black dashed line represents unity. We also plot the respective average mass-weighted and KDE-smoothed temperature and Fe abundance distributions, binned w.r.t. the mode of the central gas distribution (right panels) -- this allows us to better illustrate changes in their shape. Our results indicate that the systematic underestimation of the central ICM properties is primarily driven by projection effects, which, by introducing cooler and less abundant gas in the LoS, bias low both the temperature and Fe abundance inferences.}
    \label{fig:fov}
\end{figure*}

\begin{figure*}
    \centering
    \includegraphics[width=\linewidth]{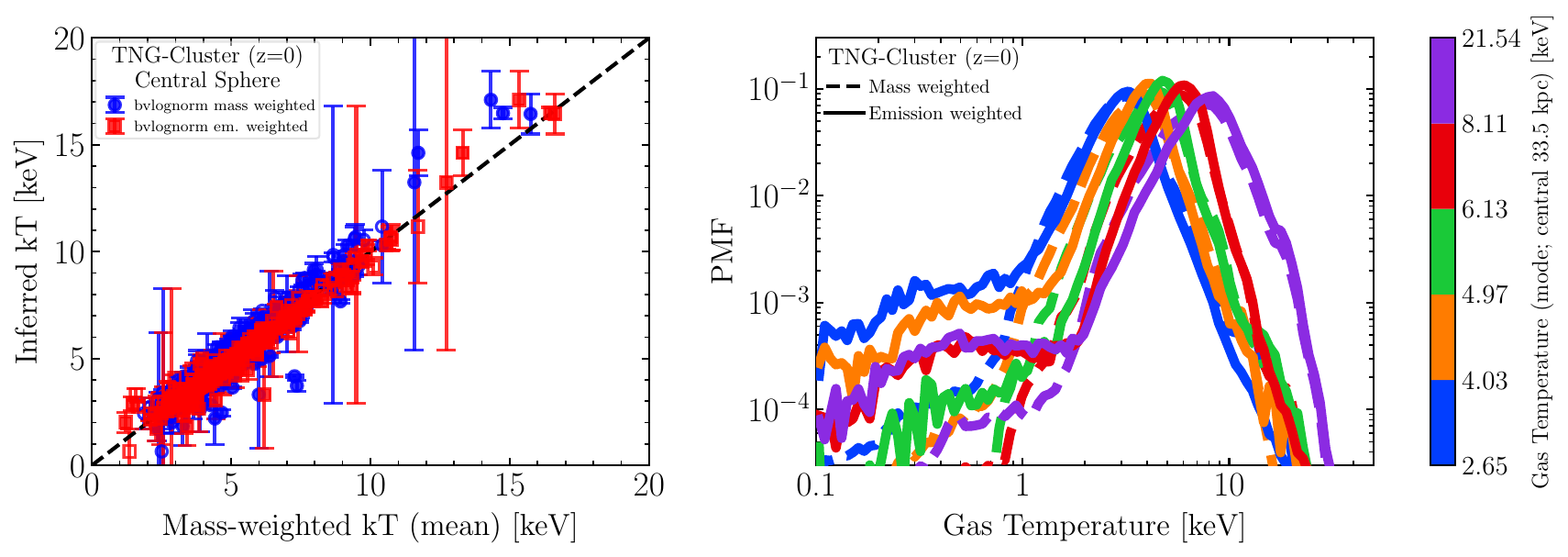}
    \includegraphics[width=\linewidth]{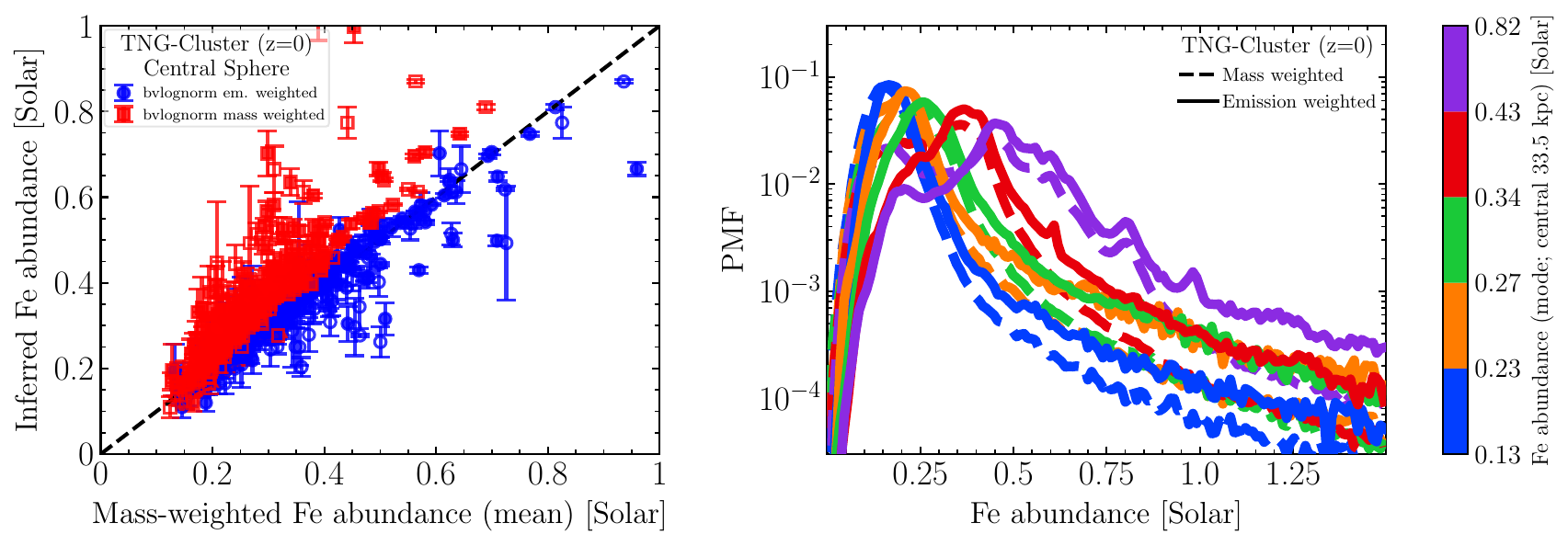}
    \caption{\textbf{Systematics due to the brightness bias in the temperature (top) and Fe abundance (bottom) distributions.} We quantify the impact of the brightness bias on the observational inferences by comparing the \texttt{bvlognorm} best-fit results to the true mass- (blue circles) and emission-weighted means (red squares) of the gas contributing to a central pointing with a \textit{XRISM}-like FoV. In the right panels, we also plot the respective average mass-weighted (dashed) and emission-weighted (solid) KDE-smoothed temperature and Fe abundance distributions, binned by the mode of the central gas distribution. Our results indicate that the emission-weighted quantities overestimate the relative contribution of the sub-keV and super-Solar gas close to the cluster centre, in turn biasing high the average Fe abundance but with minimal influence on the mean temperature.}
    \label{fig:ew}
\end{figure*}

\subsubsection{Projection effects}\label{sec:proj_eff}
Due to inherent limitations of observations, it is not possible to differentiate emission based on the 3D position of its source. Therefore, a spectrum from a central pointing contains both the contribution from the centrally-located gas and of all the gas projected along the LoS. Previous works using simulated clusters have shown that projection effects can lead to severe misinterpretations of the intrinsic thermal and chemical properties of the ICM due to the gas along the LoS \citep[e.g.][]{Zhang24,Stofanova24}.

In this section we attempt to quantify the impact of the projected gas in our inference errors by comparing the mass-weighted distributions of all the cluster gas in a central pointing with a \textit{XRISM}-like FoV to the mass-weighted distribution of the gas cells that actually reside in the region of interest, in this case a sphere of a 33.5~kpc radius located in the centre of each cluster.

In \autoref{fig:fov}, we compare the temperature distribution of the gas in the centre of clusters to the total gas along the LoS observed in our mocks (top right), and their corresponding inference errors (top left). Since the temperature distribution of the gas in the clusters centre can be in most cases approximated by a unimodal distribution, we bin the clusters in percentiles based on the mode of their temperature distribution. In fact, due to the tight correlation between ICM temperature and the halo mass, the temperature bins of \autoref{fig:fov} top right are comparable to mass bins. Differently than e.g. in \autoref{fig:Temp_inf}, here we isolate the effects of projections alone, by comparing directly mass-weighted vs. mass-weighted quantities rather than mass-weighted vs. emission weighted ones.

We see that the projection of gas from larger radii along the LoS shifts the temperature distribution towards lower temperatures (dashed vs. solid curves), due the projection of cooler gas from the outskirts of the cluster. The effect is more pronounced for systems with higher ICM temperature, or equivalently higher halo mass, due to the stronger temperature gradient up to $R_{\text{500c}}$ in those systems. This is consistent with the strong dependence of the temperature inference on halo mass, found in \autoref{fig:Temp_inf}. By comparing the difference between the inferred temperature with the \texttt{bvlognorm} model on the one side and the the mean, mass-weighted temperature of the central vs. LoS gas, we find much larger inference errors in the former case, with average offsets of -1.21~keV (18 per cent) and 0.15~keV (3 per cent), respectively, with a strong dependence on halo mass. Given the significant reduction in inference errors between the two, we can conclude that projection effects dominate the observational bias. 

Due to the projection of cooler gas, the width of the distribution systematically increases for all systems, but the effect shows a similar positive correlation with halo mass, being more pronounced for the most massive systems. The residuals on the inferred temperature width decrease by a factor of 4, from an average of 1.05 for the central gas to 0.26 for all the gas along the LoS. This suggests that the \texttt{bvlognorm} model is more representative of the distribution of temperatures along the LoS rather than the level of multi-phase gas in the central regions. However, given the inference errors, the statistical significance of the residuals effectively remains, on average, within 1~$\sigma$.

Projection effects are far more influential on the chemical structure of the ICM. In \autoref{fig:fov} (bottom) we show the mass-weighted Fe abundance distributions of both central and projected gas. Similar to the temperature distributions, we bin the clusters based on the mode of the central gas, as it consists of one prominent peak. However, unlike the temperature of the ICM, TNG-Cluster shows no clear correlation between halo mass and metal enrichment. Our results indicate that, due to projection effects, a multitude of additional abundance peaks appear in the mass-weighted Fe abundance distribution of the gas across the LoS. These peaks consist of less metal-enriched gas and are more prominent in the systems with a higher core enhancement, since for the less enriched clusters the Fe abundance distribution remains effectively unimodal. The effect arises from large amounts of less-enriched gas that is located at larger galactocentric radii intersecting the LoS, which hence biases the mean Fe abundance low. Namely, projection effects change an average inference errors between the \texttt{bvlognorm} inferred metallicity and the intrinsic mass-weighted mean Fe abundance from -0.12~Solar (22 per cent) for the central gas to 0.08~Solar (27 per cent) for the full column. The effect has a strong dependence on the mean Fe abundance of the ICM, but it is independent of halo mass, explaining the trend we see in \autoref{fig:Fe_inf}.

Overall, our results indicate that systematic errors in the inferences of the temperature and metallicity of the ICM at the centre of clusters can be greatly reduced by de-projecting the observational results, irrespective of the adopted spectral-emission model. However, we note that such an approach would require observations and analysis of spectra from the full extent of the cluster, which can not be performed within the purview of a single pointing. Still, the full minimization of projection effects may still be hampered by the assumptions on e.g. the temperature distribution of the ICM along the LoS and the geometry of the cluster.

\subsubsection{Effects of mass vs. emission weighting}

Previous studies have shown that observed ICM values are more related to emission-weighted properties than to e.g. mass-weighted properties \citep[e.g.][]{Biffi17,Biffi18}. An emission spectrum, and models to fit emission spectra, are necessarily sensitive in a weighted way towards the most luminous gas, rather than the entire mass of the ICM. As a result, physical properties of the ICM inferred by X-ray spectroscopy, such as the amount of Fe in the ICM, can be biased by the properties of the brightest gas.

In this section, we attempt to quantify how the brightest gas along the LoS contributes to the inference errors on the properties of the gas by comparing the mass- and emission-weighted distributions of the gas along the LoS, and not just that of the central sphere, as well as how it influences inferences by the \texttt{bvlognorm} model. 

Starting with the temperature, \autoref{fig:ew} shows the difference between emission- and mass-weighted distributions (top right). Overall, these are small, and primarily visible in the sub-keV tails. This is expected as the associated gas is located close to the cluster centre, is densest, and is therefore brighter than most other gas along the line of sight. The influence on the quality of the \texttt{bvlognorm} inference is similarly mild (top left). We find a shift from an average inference error of 0.15~keV (3 per cent) to -0.08~keV (1 per cent) from the mass-weighted to the emission-weighted ICM effective temperature.

The effect on the the inference of the width of the temperature distribution due to this {\it brightness bias} is relatively small, with the residuals increasing to an average of 0.4, from 0.26 for the mass-weighted case. This further highlights the previously-found trend of the Log-Normal distribution remaining unaffected by the enhancement of the sub-keV gas. (not shown)

Finally, in \autoref{fig:ew} (bottom) we examine the effects of the brightness bias on the Fe abundance distribution (bottom right). Similar to the temperatures, by favouring the brightest gas along the line of sight, there is an enhanced contribution from the super-Solar tails, that are associated with the cluster centre, and a relatively suppressed contribution from gas in the outskirts of the cluster. As a result, the emission-weighted mean Fe abundance is biased high, resulting in an average inference error of 0.08~Solar (27 per cent) compared to the average 0~Solar (<1 per cent), inferred by the \texttt{bvlognorm} model for the mass-weighted case. 

Overall, for central pointings, the brightness bias has a minimal influence on the accuracy of the observational inferences on the properties of the projected gas. This could become more a significant issue for off-centre pointings, where the projected gas becomes comparable to the medium that is targeted. Examining and quantifying the influence of those biases in such cases is left for a future study.

\begin{figure*}
\centering
        \includegraphics[width=\linewidth]{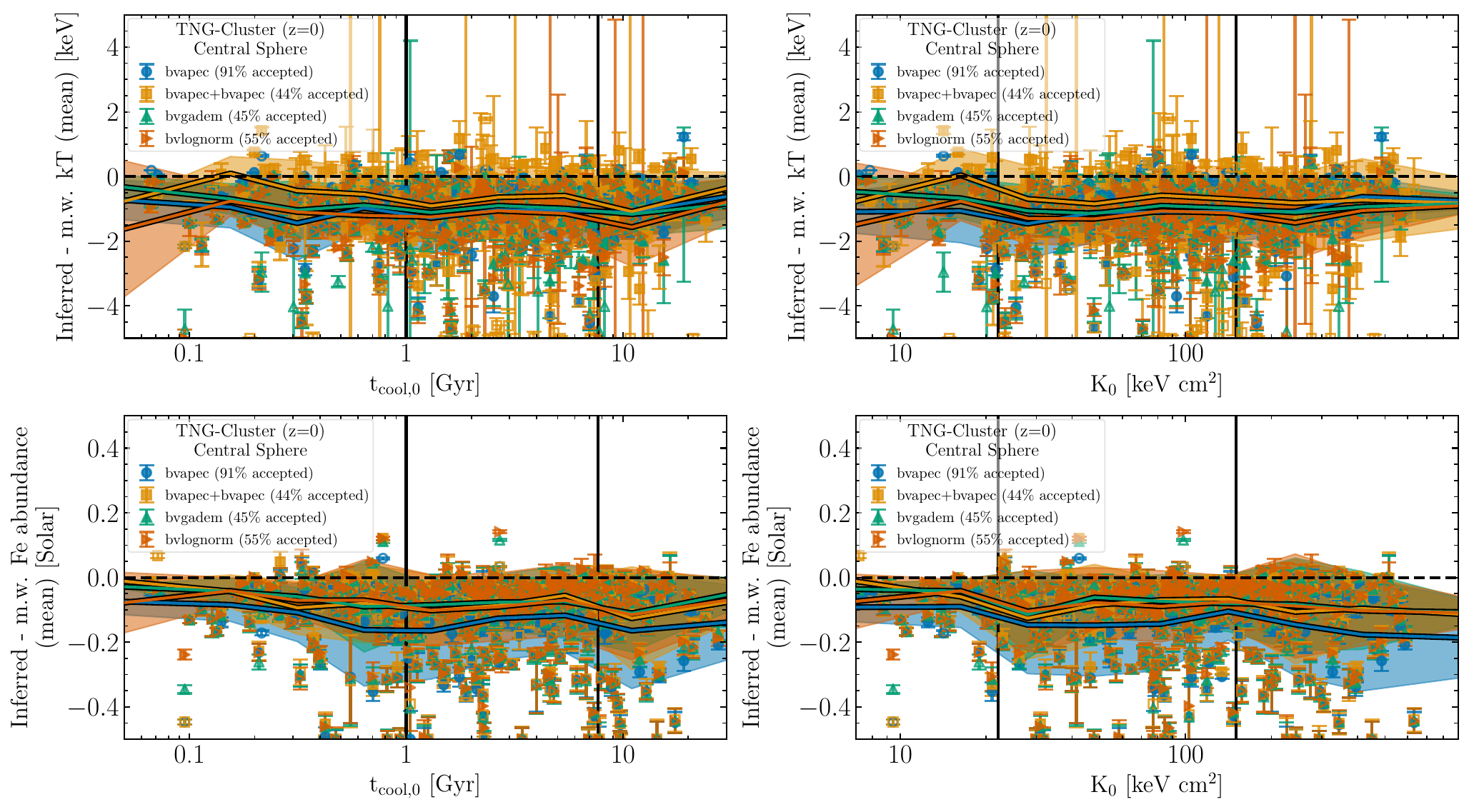}
    \caption{\textbf{Observational inferences of the properties of the ICM at the centre of clusters as a function of cool-core criteria properties.} In the top (bottom), we present the offsets between inferred temperature (Fe abundance) and the true mass-weighted mean of the central gas as a function of the cooling time (left) and central entropy (right) inside the 0.012~R$_{\text{500c}}$. The horizontal dashed line represents unity, while the vertical thick solid lines indicate strong cool-core (SCC) vs. weak cool-core (WCC) vs. non-cool-core (NCC) thresholds (see text). Solid lines and shaded regions represent the running average and 1$\sigma$ standard deviation of the difference in bins of log$_{10}(M_{\rm{500c}}/M_\odot)$=0.25, accounting only for the accepted fits. We see no clear trends between the inference errors and the the clusters' core properties, even if the cluster-to-cluster variations are smaller for strong CCs.}
    \label{fig:cc_inf}
\end{figure*}

\begin{figure*}
\centering
        \includegraphics[width=0.75\textwidth]{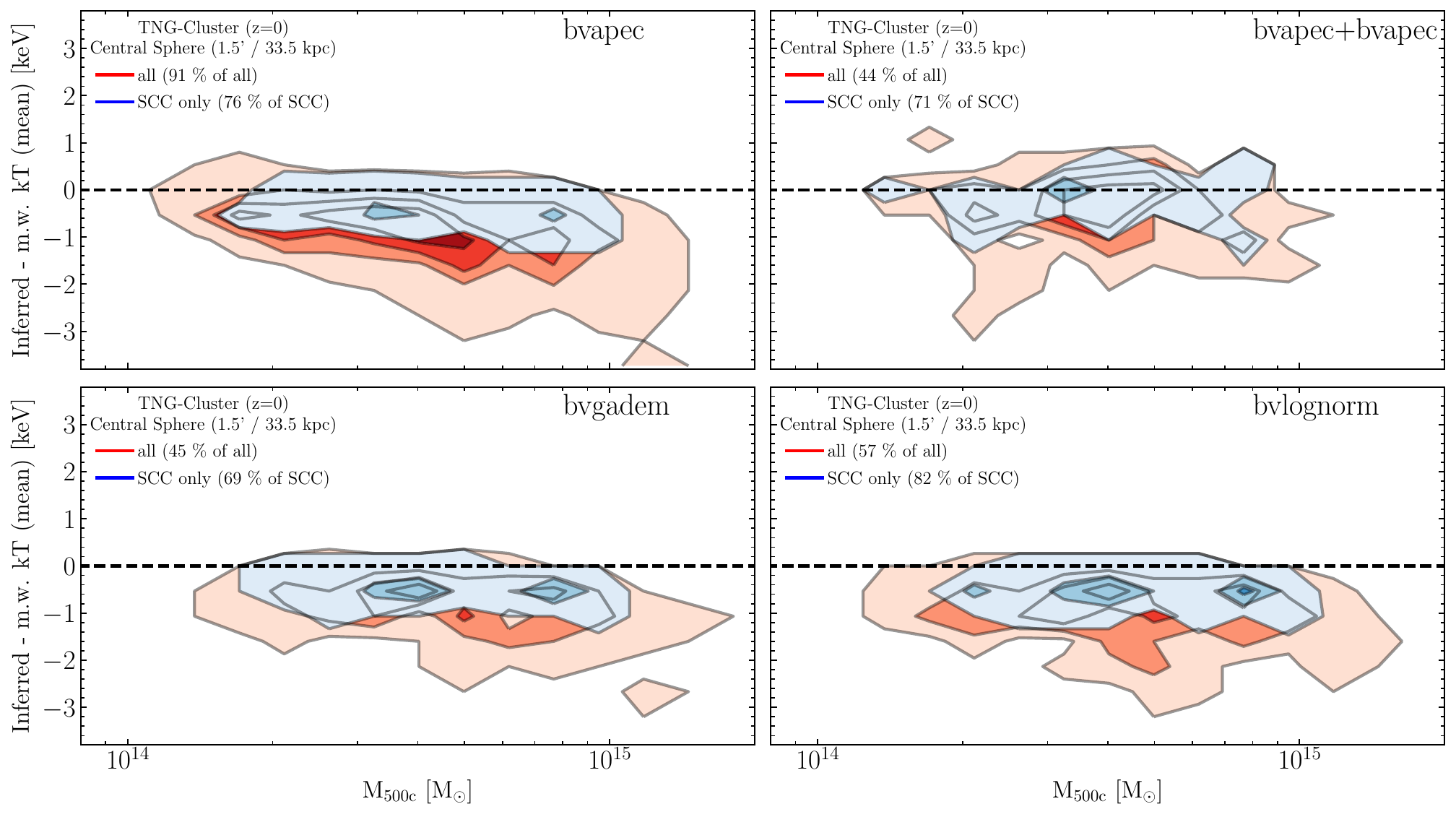}
        \includegraphics[width=0.75\textwidth]{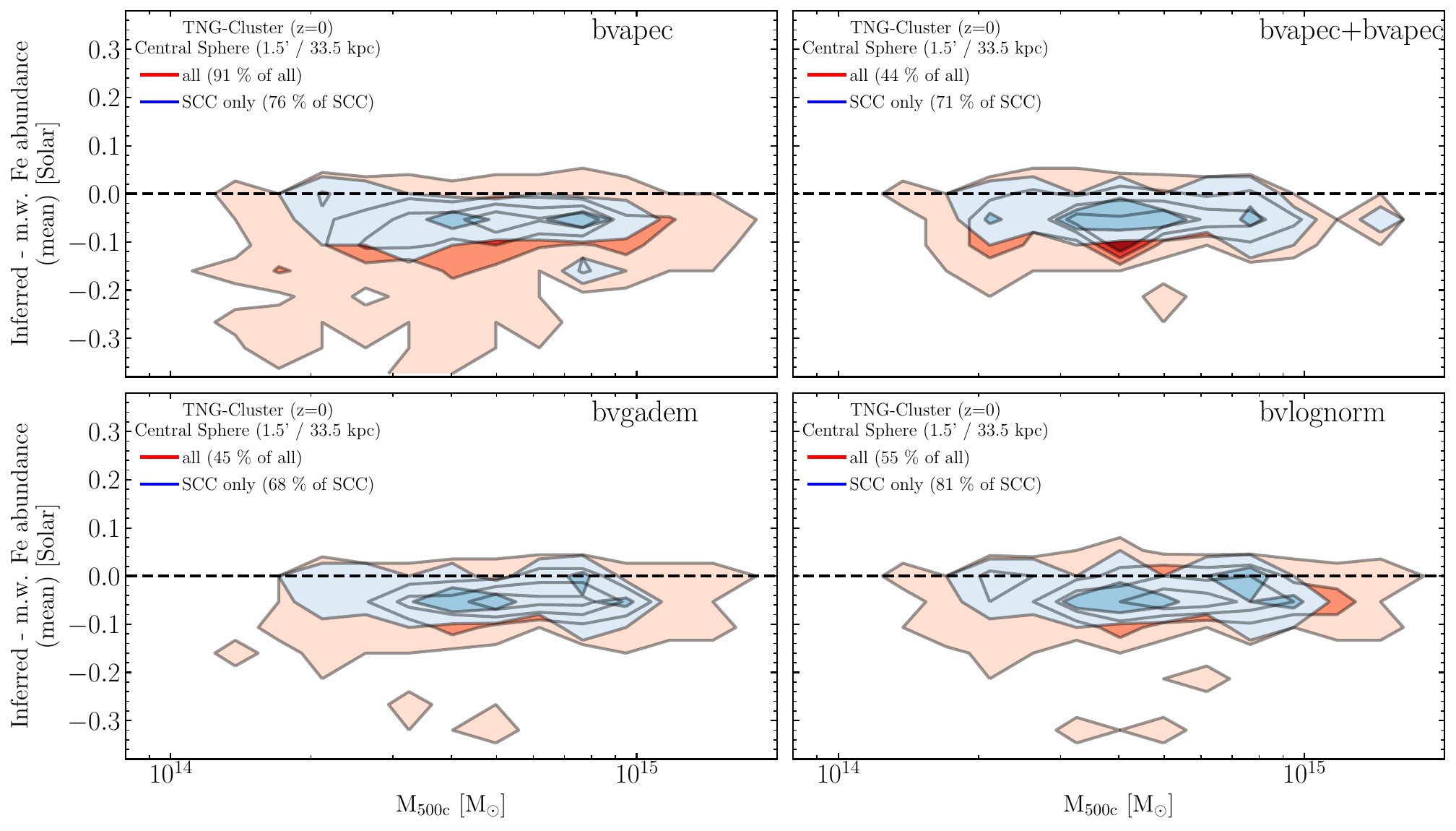}
    \caption{\textbf{Are there classes of clusters, such as cool-core clusters, for which the observational inference errors of the central ICM properties are smaller?} We focus on ICM effective temperature (top) and the Fe abundance (bottom) of the central gas and compare the  inference errors as a function of total cluster mass for all TNG-Cluster systems (orange shaded areas and contours) to those of only strong cool cores (blue shaded areas and contours; see text for details). For the former, the results are identical to those shown in \autoref{fig:Temp_inf} and \autoref{fig:Fe_inf}. Our results show that selecting only for SCCs in our sample does not change the average systematic offsets, but does remove the majority of outliers.}
    \label{fig:scc}
\end{figure*}

\subsection{Selection effects: i.e. the case of cool-core clusters}\label{sec:scc}

In our analysis, we have been considering the entire sample of 352 clusters in TNG-Cluster at $z=0$. However, samples of observed clusters tend to not be as diverse, due to selection effects. For example, many of the clusters in the literature are cool core (CC), as they are significantly brighter than their non cool-core (NCC) counterparts and considered the best targets for the study of AGN feedback. Therefore, we now investigate if the inferences of our models depend on the cool-core state of a cluster.

In \autoref{fig:cc_inf} we present the offsets between inferred and mean, mass-weighted temperature and Fe abundance for the gas in the centres of our clusters, as a function of their central cooling time and central entropy \citep[from][]{Lehle24}. For all cases, we find no strong correlation between the inference error and the cool core metric, although the scatter is larger for weak cool core (WCC) and NCC clusters.

Regarding the rejected best-fits, while a number of strong CC (SCC) clusters fail to produce an accepted fit for all models, the rejected fits disproportionally arise from the NCC subsample. Namely, $\sim$70 per cent of the SCC subsample produces an accepted double \texttt{bvapec} and \texttt{bvgadem} best-fit and 80 per cent a valid \texttt{bvapec} and \texttt{bvlognorm}; these figures are much higher than the general population acceptance rates. We note that, accounting only for SCC, the \texttt{bvlognorm} model has the best acceptance rate, surpassing even the single temperature \texttt{bvapec} one. As previously shown, constraining the temperature distribution depends on the number of counts and therefore it can be more easily accomplished the brighter the associated gas is.

Finally, we examine the associated observational inferences on the properties of the ICM, if our sample of clusters consists only of SCC clusters, that are easier to detect, compared to their NCC counterparts, due to their apparent X-ray brightness \citep[e.g.][]{Eckert11}. Selecting only for the systems with central cooling times below 1~Gyr \citep[following][]{Lehle24}, we examine their impact on temperature and Fe abundance.

In \autoref{fig:scc} we show the distribution of the CC clusters versus the full sample (shown in Figures \ref{fig:Temp_inf} and \ref{fig:Fe_inf}, for comparison). For both the temperature and Fe abundance, the average inference error increases or remains similar to that of the entire sample, respectively. For example, for the accepted \texttt{bvlognorm} fits, the change is from -1.21~keV (18 per cent) to -0.68~keV (14 per cent) and from -0.12~Solar (22 per cent) to -0.05~Solar (9 per cent) for the mean, mass-weighted temperature and Fe abundance in the central sphere, respectively. However, excluding the faintest objects does remove the majority of the outliers, suggesting that the least accurate inferences occur for NCC clusters.


\section{Summary and conclusions}\label{sec:conclusions}

In this paper, we have used the outcome of TNG-Cluster, a cosmological magnetohydrodynamical simulation of 352 massive clusters and their galaxies, to assess the accuracy with which it will be possible to infer the temperature and metallicity of clusters' centres via high-resolution X-ray spectroscopy. 

To this end, we have quantified the thermal and chemical properties of the ICM at the centre of TNG-Cluster systems at $z=0$ and we have developed an end-to-end pipeline to generate and analyse synthetic X-ray spectra that could be obtained with \textit{XRISM}/Resolve-like observations targeting the core regions of clusters at $z\sim0$. We have hence quantified the ability of commonly-used spectral-emission models (\texttt{bvapec}, double \texttt{bvapec}, \texttt{bvgadem}, and \texttt{bvlognorm}) to recover the intrinsic temperature distribution and Fe abundance of the gas. 

Our main results are as follows:

\begin{itemize}
\item[--] TNG-Cluster predicts broad distributions for both temperature and Fe abundance for the gas in the centres of clusters, with a large cluster-to-cluster variation even across systems with similar total halo mass (Section~\ref{sec:intrinsic}, \autoref{fig:gas_prop}).

\item[--] The intrinsic temperature distribution of the gas in the centre of clusters, both mass-weighted and emission-weighted, can not be completely captured by either a Normal or a Log-Normal distribution. More complicated functional forms may be necessary to accurately describe the temperature distribution of the ICM in the non-negligible fraction of systems that exhibit important tails towards both cooler and super-virial temperatures (Section~\ref{sec:intrinsic}, \autoref{fig:gas_prop_fit} and \autoref{fig:best_fit_val}).

\item[--] The central ICM temperature inferred via all spectral-emission models is consistent with the true emission-weighted average along the line of sight, but systematically underestimates the mass-weighted temperature of the gas that is truly located in the clusters centre. This inference bias reads, on average (i.e. across the whole population of 352 clusters), -1.2 keV (17 per cent; -0.73,+0.72 keV for the 16th-84th percentiles), so it is arguably small, but has a dependence on halo mass (Section~\ref{sec:temp}, \autoref{fig:Temp_inf}). For $10^{15}~\rm{M}_\odot$ clusters, all spectral-emission models return mean temperatures that are smaller by 1.5 keV, on average, than the true mass-weighted average one.

\item[--] The width of the temperature distribution inferred by the \texttt{bvlognorm} model, is, on average, overestimated compared to the true mass-weighted distribution of the central ICM, but within 1$\sigma$ of the emission-weighted distributions. In contrast, the \texttt{bvgadem} model overestimates the width of both mass- and emission-weighted temperature distributions to a larger extent (Section~\ref{sec:temp}, \autoref{fig:width_inf}).

\item[--] The inferred Fe abundance is consistent with the true mean, emission-weighted Fe abundance of the gas along the line of sight, i.e. for a central pointing with a \textit{XRISM}-like FoV. However, all models similarly underestimate the mean Fe abundance in the cluster centre by about 0.12~Solar (22 per cent), which is significantly larger than the statistical uncertainties usually associated to the measurements (Section~\ref{sec:Fe_abund}, \autoref{fig:Fe_inf}).

\item[--]  The rejection of fits of the X-ray spectra correlates with the number of collected photon counts but, interestingly, not with errors in the inference. More than half of the initially rejected \texttt{bvlognorm} fits do succeed with longer exposure times. However, metal-poor systems and systems with temperature distributions that deviate strongly from a Log-Normal, fail to produce an accepted spectral fit, even if we account for longer exposures (300ks vs 100ks; Section~\ref{sec:insights}, \autoref{fig:counts}).

\item[--] Projection effects are the primary reason of the offset between inferred and the actual properties of the central ICM. In contrast, the brightness bias (emission- vs. mass -weighted) has a smaller impact, at least for central pointings (Section~\ref{sec:obs_bias}, \autoref{fig:fov}, \autoref{fig:ew}).

\item[--] When considering only strong cool-core clusters, as is the case in many observational samples, model inferences on central ICM temperature improve only marginally for the average cluster (with e.g. inference errors of 0.7 keV on average compared to 1.2 keV across a randomly-selected overall sample). The CC-selection effect on the inference of Fe abundance is even smaller. The main advantage of selecting for SCCs is the removal of strong outliers, i.e. of clusters for which the observational inferences of ICM properties are catastrophically wrong (Section~\ref{sec:scc}, \autoref{fig:scc}).
\end{itemize}

Overall, our analysis highlights the importance of properly modelling the complex temperature and metallicity distributions of the ICM at the centre of clusters. The different spectral-emission models that we have considered have different strengths and weaknesses, or practical justifications. Barring the double \texttt{bvapec}, which returns highly inaccurate results for more clusters than the other models, all studied assumptions return overall similarly accurate inferences of the effective temperature and average metallicity of the ICM at the centre of clusters.  However, based on the predictions of TNG-Cluster, we recommend to use models that allow for a temperature distribution, as both single and double temperature models simply fail at capturing the diversity and complexity of the ICM.  Similarly, the single-value Fe-abundance assumption, typical of all ICM X-ray spectra analyses, seems utterly inadequate to represent the multifacetedness of the ICM enrichment, even in the innermost regions of clusters -- a qualitative shift in the way ICM X-ray spectra are fitted for, and hence interpreted, seems of the essence. Finally, due to the large influence of the effects associated with observational realism, particularly projection effects, a careful deprojection to the 3D properties of the gas from observed data is required.

While we have primarily focused on the ICM Fe abundance, in future work we will assess the accuracy of \textit{XRISM}/Resolve-like observations and analyses in recovering the abundances of other metal species, in addition to gas kinematics. It remains to be quantified how the figures put forward here translate to lower, CCD-like resolution spectra, such as with \textit{XMM-Newton}, to observations targeting off-centre pointings and cluster outskirts, and to science cases that rely on spatially-resolved analyses.


\section*{Data Availability}

All the output of the TNG-Cluster simulation is now publicly available, as are the IllustrisTNG simulations: see \url{www.tng-project.org/data} and \citet{Nelson19}. The data generated in this project, including {\textit XRISM} Resolve mocks, spectra, and spectral-fitting results, will be made available upon acceptance. 

\section*{Acknowledgements}

We thank the reviewer for their constructive and useful suggestions. DC and AP thank Joey Braspenning for useful comments on the paper. DC is a fellow of the International Max Planck Research School for Astronomy and Cosmic Physics at the University of Heidelberg (IMPRS-HD) and, together with AP, acknowledges funding from the European Union (ERC, COSMIC-KEY, 101087822, PI: Pillepich). DN acknowledges funding from the Deutsche Forschungsgemeinschaft (DFG) through an Emmy Noether Research Group (grant number NE 2441/1-1). 

The TNG-Cluster simulation suite has been executed on several machines: with compute time awarded under the TNG-Cluster project on the HoreKa supercomputer, funded by the Ministry of Science, Research and the Arts Baden-Württemberg and by the Federal Ministry of Education and Research; the bwForCluster Helix supercomputer, supported by the state of Baden-Württemberg through bwHPC and the German Research Foundation (DFG) through grant INST 35/1597-1 FUGG; the Vera cluster of the Max Planck Institute for Astronomy (MPIA), as well as the Cobra and Raven clusters, all three operated by the Max Planck Computational Data Facility (MPCDF); and the BinAC cluster, supported by the High Performance and Cloud Computing Group at the Zentrum für Datenverarbeitung of the University of Tübingen, the state of Baden-Württemberg through bwHPC and the German Research Foundation (DFG) through grant no INST 37/935-1 FUGG. The material is based upon work supported by NASA under award number 80GSFC24M0006. 

All the analysis and computations associated to this paper have been realized on the Vera cluster of the MPCDF.

\bibliographystyle{mnras}
\bibliography{ms}


\appendix

\section{spectral-emission model systematic uncertainties}\label{sec:systematics}
This section reports in detail, in a series of Tables (\ref{tab:errors_1}, \ref{tab:errors_2}), the systematic uncertainties, associated with the choice of spectral-emission model, on the inference of the true properties of the ICM, as predicted by TNG-Cluster. 

\begin{table*}
\begin{tabularx}{\linewidth} { 
  >{\hsize=0.25\linewidth\raggedright\arraybackslash}X 
  >{\hsize=0.2\linewidth\centering\arraybackslash}X 
  >{\hsize=0.2\linewidth\centering\arraybackslash}X 
  >{\hsize=0.2\linewidth\centering\arraybackslash}X 
  >{\hsize=0.2\linewidth\centering\arraybackslash}X }
  \hline \hline
    Inferred - True value& \texttt{bvapec} & double \texttt{bvapec} & \texttt{bvgadem} & \texttt{bvlognorm} \\
    \hline \\
    \multicolumn{5}{c}{Effective Temperature (kT) [keV]} \\ \\
    \hline \\
    Mass-weighted central gas (all) & -1.05 (-0.75,+0.74) & -1.32 (-1.07,+1.22) & -1.18 (-0.78,+0.74) & -1.21 (-0.73,+0.72)\\
    Mass-weighted central gas (accepted only) & -1.09 (-0.72,+0.72) & -0.68 (-0.92,+0.87) & -0.98 (-0.59,+0.54) & -1.17 (-0.67,+0.68)\\
   emission-weighted FoV gas (all) & 0.09 (-0.24,+0.16) & -0.17 (-0.58,+0.86) & -0.05 (-0.17,+0.22) & -0.08 (-0.22,+0.24)\\
   emission-weighted FoV (accepted only) & 0.06 (-0.21,+0.15) & 0.22 (-0.95,+0.69) & 0.00 (-0.16,+0.11) & -0.15 (-0.17,+0.14) \\
    \hline \\
    \multicolumn{5}{c}{Temperature width Residuals ($\sigma_{\rm{kT,res}}$) (\ref{eq:sigma_res})} \\ \\
    \hline \\
    Mass-weighted central gas (all) & & & 1.56 (-1.77,+0.55) & 1.05 (-1.28,+0.70) \\
    Mass-weighted central gas (accepted only) & & & 2.08 (-1.73,+0.64) & 1.51 (-1.21,+0.79) \\
    emission-weighted FoV gas (all) & & & 1.27 (-1.67,+0.56) & 0.41 (-1.05,+0.85) \\ 
    emission-weighted FoV gas (accepted only) & & & 1.83 (-1.58,+0.81) & 0.81 (-0.89,+0.92) \\
    \hline \\
    \multicolumn{5}{c}{Fe abundance [Solar]} \\ \\
    \hline \\
    Mass-weighted central gas (all) & -0.13 (-0.11,+0.10) & -0.12 (-0.11,+0.09) & -0.12 (-0.12,+0.09) & -0.12 (-0.12,+0.09)  \\
    Mass-weighted central gas (accepted only) & -0.14 (-0.11,+0.10) & -0.09 (-0.05,+0.07) & -0.08 (-0.04,+0.06) & -0.08 (-0.04,+0.07) \\
    emission-weighted FoV gas (all) & -0.02 (-0.02,+0.03) & -0.01 (-0.02,+0.03) & -0.01 (-0.02,+0.03) & -0.00 (-0.02,+0.03) \\
    emission-weighted FoV gas (accepted only) & -0.02 (-0.02,+0.03) & 0.01 (-0.01,+0.03) & 0.01 (-0.01,+0.03) & -0.01 (-0.01,+0.03) \\
    \hline \hline
    \end{tabularx}
    \caption{\textbf{Absolute inference errors, i.e. differences between inferred and true values, from each spectral-emission model's inference.} We report the mean difference between inferred and true value for the models used in this work. The 16$^{\rm{th}}$ and 84$^{\rm{th}}$ percentiles are reported with respect to the mean in parenthesis}    
    \label{tab:errors_1}
\end{table*}

\begin{table*}
\begin{tabularx}{\linewidth} { 
  >{\hsize=0.25\linewidth\raggedright\arraybackslash}X 
  >{\hsize=0.2\linewidth\centering\arraybackslash}X 
  >{\hsize=0.2\linewidth\centering\arraybackslash}X 
  >{\hsize=0.2\linewidth\centering\arraybackslash}X 
  >{\hsize=0.2\linewidth\centering\arraybackslash}X }
  \hline \hline
    Inferred / True value& \texttt{bvapec} & double \texttt{bvapec} & \texttt{bvgadem} & \texttt{bvlognorm} \\
    \hline \\
    \multicolumn{5}{c}{Effective Temperature (kT) [keV]} \\ \\
    \hline \\
    Mass-weighted central gas (all) & -0.14 (-0.08,+0.06) & -0.19 (-0.12,+0.17) & -0.18 (-0.06,+0.09) & -0.18 (-0.08,+0.07) \\
    Mass-weighted central gas (accepted only) & -0.15 (-0.07,+0.06) & -0.11 (-0.18,+0.16) & -0.15 (-0.06,+0.06) &-0.19 (-0.08,+0.08) \\
    emission-weighted FoV gas (all) & 0.05 (-0.07,+0.02) & -0.02 (-0.14,+0.20) & -0.01 (-0.03,+0.04) & -0.01 (-0.05,+0.04) \\
    emission-weighted FoV gas (accepted only) & 0.04 (-0.06,+0.01) & 0.06 (-0.23,+0.17) & 0.00 (-0.03,+0.02) & -0.03 (-0.04,+0.03) \\
    \hline \\
    \multicolumn{5}{c}{Fe abundance [Solar]} \\ \\
    \hline \\
    Mass-weighted central gas (all) & -0.25 (-0.19,+0.17) & -0.22 (-0.20,+0.20) & -0.23 (-0.20,+0.20) & -0.22 (-0.20,+0.17) \\
    Mass-weighted central gas (accepted only) & -0.26 (-0.19,+0.16) & -0.16 (-0.13,+0.13) & -0.15 (-0.11,+0.10) & -0.15 (-0.12,+0.11) \\
    emission-weighted FoV gas (all) & -0.04 (-0.08,+0.04) & 0.00 (-0.09,+0.09) & -0.01 (-0.09,+0.09) & 0.01 (-0.08,+0.10) \\
    emission-weighted FoV gas (accepted only) & -0.03 (-0.07,+0.07) & 0.03 (-0.05,+0.07) & 0.03 (-0.04,+0.06) & 0.05 (-0.06,+0.07) \\
    \hline \hline
    \end{tabularx}
    \caption{\textbf{Relative inference errors ((inferred-true)/true), of the studied ICM properties throughout this paper.} Similar to \autoref{tab:errors_1} bit for the ratio of inferred to true value instead.}   
    \label{tab:errors_2}
\end{table*}

\section{Inferred LoS velocity dispersion}\label{sec:sigma}
\begin{figure}
    \centering
    \includegraphics[width=\columnwidth]{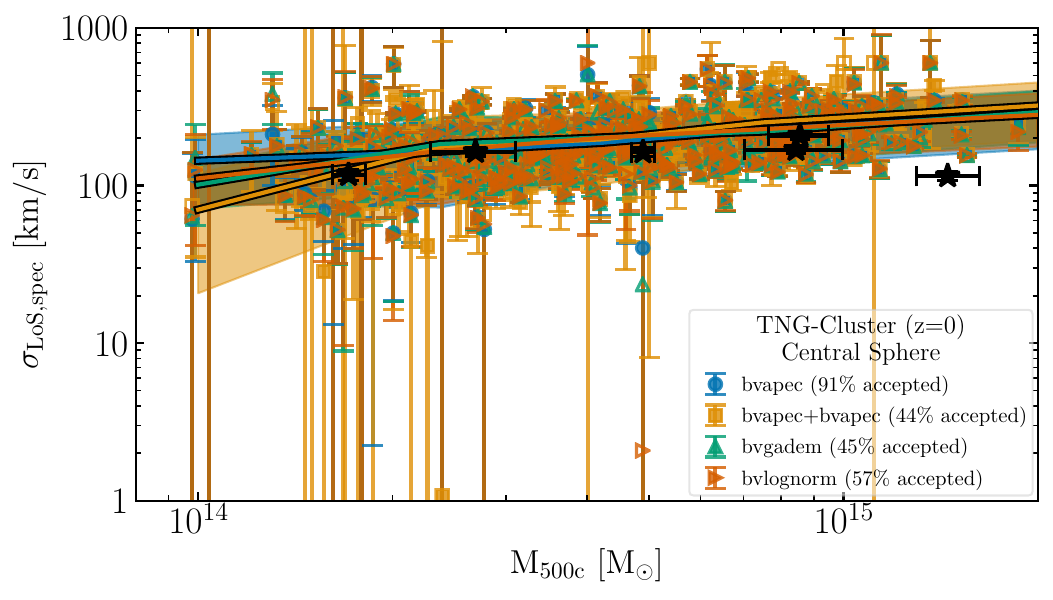}
    \caption{\textbf{LoS velocity dispersions of the ICM at the centre of clusters, from TNG-Cluster and from \textit{XRISM} observational analyses.} For the same spectral fits and modelling of \autoref{fig:Temp_inf}, we showcase the inferred velocity dispersion by all models used in this work, for a central pointing with a \textit{XRISM}-like FoV. Annotations are as in \autoref{fig:Temp_inf} and, as there, we indicate the running average and 1$\sigma$ standard deviation of the accepted fits as a function of each cluster's M$_{\rm{500c}}$. We also include the observationally-derived velocity dispersion values from \textit{XRISM}/Resolve observations of the cores of Centaurus, Hydra A, Perseus, Abell 2029, Coma and Ophiuchus, in order of increasing halo mass. We highlight that all spectral-emission models infer similar velocity dispersions, suggesting that the inference of the gas kinematics is somewhat independent of the modelling of the ICM's temperature distribution, at least for the average cluster. If observed and simulated clusters can be compared without accounting for selection biases (as done here), it would seem that predictions from TNG-Cluster are overall consistent with observations, since all of them fall within 1$\sigma$ from our accepted fits, population averages.}
    \label{fig:vel_disp}
\end{figure}
In this section, we report, in addition to the temperature and metallicity properties of the ICM -- the focus of the paper --, also the LoS velocity dispersion values inferred according to each spectral-emission model. We do so for the full sample of studied clusters. As previously mentioned, comparing these values to the intrinsic 3D velocity dispersion of the gas cells from the simulation exceeds the scope of this work. However, considering potential biases due to degeneracies among all free parameters and considering the current interest in the kinematics of cluster centres enables by \textit{XRISM}, we want to provide results also on our kinematic inferences.

In \autoref{fig:vel_disp} we present the inferred velocity dispersions for all spectral-emission models used in this work as a function of redshift. We note that we find little to no difference in the inferred values between models, with the possible exception of the double \texttt{bvapec} model for low mass halos. This suggests that the inferred kinematics appear to be, for the most part or at least for the average cluster, unaffected by the parametrisation of the ICM's thermal distribution.

For comparison, we also include in \autoref{fig:vel_disp} the observationally-inferred velocity dispersions from the cores of a few selected galaxy clusters, as available from the current literature (large black markers, from lower to higher masses: Centaurus, Hydra A, Perseus, Abell 2029, Coma and Ophiuchus). We see that observations are consistent with the TNG-Cluster predictions, deviating by less than 1$\sigma$ from the average trends. The only outlier appears to be Ophiuchus, at the highest mass end. However, in the observational setup, the central region only covers the central 25~kpc, compared to the 33.5~kpc used in this work, and the velocity dispersion appears to increase with radius \citep[][]{Fujita25}. This might imply that the velocity dispersion of Ophiuchus may also fall within 1$\sigma$ of the TNG-Cluster predictions, if the central pointings were matched. This preliminary and at-face-value comparison may serve as a useful starting point for future, more tailored analyses \citep[e.g.][]{XrismCollab25c,XrismCollab25d}.

\section{On fitting the temperature distributions of simulated ICM}\label{sec:residuals}
\begin{figure*}
    \centering
    \includegraphics[width=\linewidth]{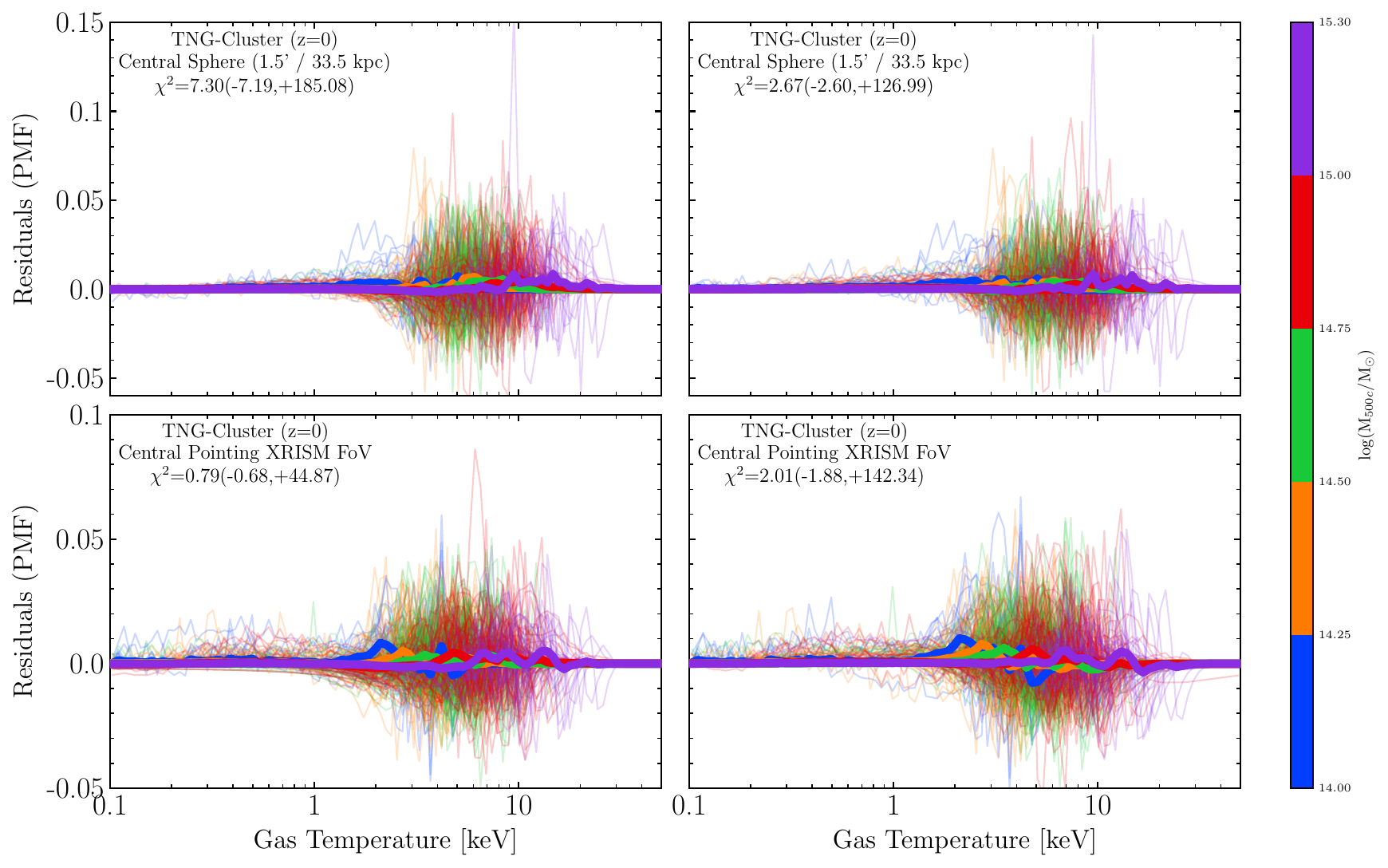}
    \caption{\textbf{Residuals from fitting the intrinsic thermal distribution, in an individual cluster basis.} The thin curves show the residuals (intrinsic-fitted) from fitting the temperature distribution of the mass-weighted gas in the 3D clusters' centres (top) and the emission-weighted gas for a central pointing with \textit{XRISM}-like FoV (bottom), with a Normal (left) and a Log-Normal distribution (right), following \autoref{fig:gas_prop_fit}. The thick curves indicate the running averages per halo mass bin, as indicated. On average, both the Normal and a Log-Normal functional forms return residuals of the order of 1 per cent, with systematic biases, as highlighted in the main text. However, the offsets can become significant for individual clusters in specific temperature ranges, which is reflected in the large spread of $\chi^2$ values around the median.}
    \label{fig:residuals}
\end{figure*}

As discussed in Section~\ref{sec:intrinsic}, we fit the intrinsic temperature distribution of the ICM predicted by TNG-Cluster with both a Normal and a Log-Normal distribution. In \autoref{fig:residuals} we present the residuals, i.e. the difference between intrinsic and modelled probability mass functions, for both fit results. On average (thick curves), the offsets do not exceed the 1 per-cent level, suggesting that both Normal and a Log-Normal functional forms can provide a reasonable description of the thermal properties of the gas across our entire cluster sample. 

However, the tensions are more significant for individual halos (thin curves), with a large cluster-to-cluster variation reflected in the large scatter of $\chi^2$ values. As we have already highlighted for the entire sample, the tensions primarily arise from highly asymmetric intrinsic distributions, due to extended tails of either hotter or cooler gas, relative to the mode of the distribution. We note that for both functional forms, we find an increase in the residuals on the sides of the temperature peak due to over-/under-fitting, as explained in the text.

\end{document}